\PassOptionsToPackage{table}{xcolor}
\documentclass{article}
\usepackage{iclr2027_conference,times}
\iclrfinalcopy % Display the supplied author list.
\usepackage{amsmath}
\usepackage{bbm}
\usepackage{times}
\usepackage{latexsym}
\usepackage{amssymb}
\usepackage{float}
\usepackage{placeins} % Keep pending result figures before the conclusion.
\usepackage{graphicx}
\usepackage{subcaption}
\usepackage{colortbl}
\usepackage{xcolor}
\usepackage{lineno}
\usepackage{array}
\usepackage{booktabs}
\usepackage{multirow}
\usepackage{adjustbox}
\usepackage{makecell}
\usepackage{enumitem}
\usepackage{tabularx}
\usepackage{pifont}
\usepackage{wrapfig}
\definecolor{tableheadcolor}{RGB}{230, 235, 255}
\renewcommand{\tabularxcolumn}[1]{m{#1}}
\newcolumntype{Y}{>{\centering\arraybackslash}X}
\usepackage[T1]{fontenc}
\usepackage[utf8]{inputenc}
\usepackage{hyperref}
\usepackage{microtype}

\usepackage{inconsolata}

\usepackage{dashrule}

\usepackage{graphicx}

\usepackage{tipa}

\usepackage{booktabs}
\usepackage[table]{xcolor}
\usepackage{tikz}

\usepackage{booktabs}   % 必须：提供三线表 \toprule, \midrule, \bottomrule
\usepackage{multirow}   % 必须：提供跨行单元格
\usepackage{graphicx}   % 必须：提供 \resizebox

\hypersetup{
  colorlinks=true,
  linkcolor=red,
  citecolor=cyan,
  filecolor=magenta,
  urlcolor=magenta,
}

\DeclareRobustCommand{\revised}[1]{{#1}}
\newenvironment{rebuttalrevision}{\begingroup}{\par\endgroup}
\definecolor{resultbandlight}{RGB}{178,218,222}
\definecolor{resultbanddark}{RGB}{126,187,190}
\newcommand{\rankfirst}[1]{\textbf{#1}}
\newcommand{\ranksecond}[1]{\underline{#1}}

\DeclareRobustCommand{\rankcircle}[1]{%
    \tikz[baseline=(rank.base)]{%
        \node[draw, circle, inner sep=0pt, outer sep=0pt,
              minimum size=1.05em, line width=0.3pt,
              font=\normalfont, text height=0.7em, text depth=0pt]
              (rank) {#1};%
    }%
}
\newcommand{\overallrank}[2]{#1\,\makebox[0.9em][l]{\raisebox{0.55ex}{\fontsize{5.2}{5.2}\selectfont\rankcircle{#2}}}}
\newcommand{\overallplain}[1]{#1\,\makebox[0.9em][l]{}}

\AtBeginDocument{%
  \setlength{\textfloatsep}{8pt}     % 页顶、页底的普通浮动体与正文
  \setlength{\dbltextfloatsep}{8pt}  % 双栏模式下跨栏浮动体与正文
  \setlength{\intextsep}{8pt}        % 插入正文中的浮动体与上下文
}

\usepackage{tcolorbox} 
\tcbuselibrary{most} 
\usepackage{fontawesome5}    % 用于那个小齿轮图标 \faCog

\vspace{-0.6em}

\title{Think2Listen: Evidence-Centric Learning for Agentic Audio Understanding with LALMs}

\author{%
\parbox[t]{\dimexpr\textwidth-2\tabcolsep\relax}{%
\raggedright\normalsize\bfseries
Jin Wang\textsuperscript{1,2}\thanks{Equal contribution.},
Liang Lin\textsuperscript{3}\footnotemark[1],
Kaiwen Luo\textsuperscript{3}, Weiliu Wang\textsuperscript{4},
Yuxuan Li\textsuperscript{5}\\[2pt]
Zengxin Han\textsuperscript{2}, Gan Zhang\textsuperscript{2},
Jingtong Wu\textsuperscript{2}, Yufei Guo\textsuperscript{6},
Kun Wang\textsuperscript{3}\thanks{Corresponding author.}\\[5pt]
\normalfont\small
\mbox{\textsuperscript{1}Xi'an University of Electronic Science and Technology}\quad
\mbox{\textsuperscript{2}Ant International}\\
\mbox{\textsuperscript{3}Nanyang Technological University}\quad
\mbox{\textsuperscript{4}Beijing University of Posts and Telecommunications}\\
\mbox{\textsuperscript{5}University of the Chinese Academy of Sciences}\quad
\mbox{\textsuperscript{6}Peking University}
}}

\definecolor{caseActor}{HTML}{2B6F94}
\definecolor{caseAudio}{HTML}{347F87}
\definecolor{caseEvidence}{HTML}{557C78}
\definecolor{caseInk}{HTML}{253238}
\definecolor{caseRule}{HTML}{CEDADC}

\newcommand{\CaseGap}{\par\vspace{3pt}}
\newcommand{\CaseRole}[2]{%
  \par\noindent{\rmfamily\bfseries\fontsize{7.4}{9}\selectfont\color{#1}#2}%
  \par\nobreak\vspace{2pt}}
\newcommand{\CaseDivider}{%
  \par\vspace{5pt}\noindent{\color{caseRule}\rule{\linewidth}{0.35pt}}%
  \par\vspace{3pt}}
\newcommand{\CaseText}[1]{%
  {\ttfamily\fontsize{9}{10.4}\selectfont\color{caseInk}\raggedright
  \hyphenpenalty=10000\exhyphenpenalty=10000 #1\par}}
\newtcolorbox{papercasebox}[3][1 ASK]{%
  enhanced,breakable,
  colframe=#2,colback=#2!2!white,colbacktitle=#2,
  coltitle=white,coltext=caseInk,
  boxrule=0.65pt,arc=1.2pt,outer arc=1.2pt,
  boxsep=0pt,left=9pt,right=9pt,top=7pt,bottom=8pt,
  lefttitle=9pt,righttitle=9pt,toptitle=4pt,bottomtitle=4pt,
  fonttitle=\bfseries\fontsize{10}{12}\selectfont,
  title={#3\hfill{\normalfont\rmfamily\fontsize{7.5}{9}\selectfont #1}},
  title after break={#3\hfill{\normalfont\fontsize{7.5}{9}\selectfont #1 / CONTINUED}},
  before upper={\setlength{\parskip}{0pt}},
  before skip=8pt,after skip=0pt,pad at break*=6pt,lines before break=5
}
\newcommand{\CaseResult}[2]{%
  \begin{tcolorbox}[colback=caseEvidence!8!white,colframe=caseEvidence!8!white,
    boxrule=0pt,arc=1pt,boxsep=0pt,left=6pt,right=6pt,top=5pt,bottom=5pt,
    before skip=7pt,after skip=0pt]
  \CaseRole{caseEvidence}{FINAL ACTION\hfill GROUND-TRUTH ANSWER}
  {\ttfamily\fontsize{9}{10.4}\selectfont\color{caseInk}%
  <answer>#1</answer>\hfill #2\par}
  \end{tcolorbox}}
\begin{document}

\maketitle
\fancyhead{} % Preserve the original blank header.

\begin{abstract}
Recent advances in large audio-language models (LALMs) have broadened
audio understanding across speech, sound, and music, with tool-based agents
further supporting these tasks by invoking predefined audio tools.
However, a persistent gap remains between perceiving acoustic content and
using it as evidence for reasoning: recognizing what is audible does not
ensure that a model can identify which cues resolve an uncertain decision.
To bridge this gap, we propose \textbf{\texttt{Think2Listen}}, an evidence-centric
learning framework for agentic audio understanding.
Guided by \textit{evidence before commitment}, \textbf{\texttt{Think2Listen}}
couples iterative reasoning with a targeted LALM interaction to actively seek 
acoustic cues that distinguish competing hypotheses and revise candidate
judgments based on the observations returned.
It learns \textit{what to ask}, \textit{how to use evidence}, and
\textit{when to stop}, using observation-induced changes in candidate-state to
inform training credit.
Extensive evaluations on two benchmarks spanning seven audio categories
demonstrate that \textbf{\texttt{Think2Listen}} achieves
\textbf{(I)} enhanced multi-step audio reasoning, with gains in
overall accuracy of 5.08\%--17.67\% over competing audio agents; and
\textbf{(II)} improved audio understanding and reasoning across speech,
sound, and music, boosting overall accuracy by
2.29\%--13.10\% over the corresponding baselines.
The source code is available at \url{https://github.com/JinWang79/Think2Listen}.
\end{abstract}

\section{Introduction}
% Paragraph 1: research background and existing technical directions; no summary.
Large Audio-Language Models (LALMs) have demonstrated increasingly broad
capabilities across speech, sound, and music, enabling diverse auditory tasks
through a shared language-based
interface~\citep{yang2025towards,luo2026survey}.
In parallel, language-model agents have demonstrated how reasoning and
environmental feedback can guide tool use and subsequent
decisions~\citep{yao2022react,shinn2023reflexion}.
In the audio domain, agentic systems coordinate language models with
specialized audio capabilities to support interactive task
solving~\citep{huang2024audiogpt,wijngaard2025audiotoolagent}.
Recent work has explored acquiring and refining task-relevant audio
information during reasoning~\citep{xiong2025thinking,rong2026audiogenie,zhang2026echo},
as well as learning when and which audio tools to
invoke~\citep{tong2026autagent,chen2026audiorouter,wang2026hear}.

% Despite these advances, a persistent gap separates acoustic perception
% from the use of evidence in reasoning. 
% Models can describe an audio scene yet fail to answer discriminative questions about it~\citep{kuan2024understanding}. 
% Even acoustic attributes recoverable from their encoders may remain poorly reflected in final predictions~\citep{sun2026sonicbench}.
% These findings suggest that the difficulty extends beyond recognizing sounds to using acoustic cues to distinguish competing interpretations. Studies of temporal reasoning support this view: how models allocate attention to audio affects their task performance~\citep{kulkarni2026closer}.
% Additional reasoning or interaction does not automatically bridge this
% gap: reasoning can elaborate an incorrect acoustic premise~\citep{tian2025step}, while intermediate evidence-gathering stages can degrade performance relative to direct listening~\citep{wang2026audio}.
% The resulting challenge is to identify the evidence needed for an unresolved decision and assess whether new observations justify revising it.

Despite these advances, a persistent gap separates acoustic perception
from the use of evidence in reasoning.
Even when acoustic information is available, models may struggle to
determine which cues are relevant to the question and how they support
or rule out competing interpretations.
Models can describe an audio scene yet fail to answer discriminative
questions about it~\citep{kuan2024understanding}.
Even acoustic attributes recoverable from their encoders may remain
poorly reflected in final predictions~\citep{sun2026sonicbench}.
Together, these findings motivate treating acoustic cues as evidence for comparing candidate answers in agentic audio understanding.

% Despite these advances, a persistent gap separates acoustic perception
% from the use of evidence in reasoning.
% Models can describe an audio scene yet fail to answer discriminative
% questions about it~\citep{kuan2024understanding}.
% Even acoustic attributes recoverable from their encoders may remain poorly
% reflected in final predictions~\citep{sun2026sonicbench}.
% Studies of temporal reasoning further show that
% redistributing attention across audio tokens can be more effective than
% increasing total audio attention~\citep{kulkarni2026closer}.
% These findings suggest that recognizing acoustic content does not ensure
% its effective use in a decision. Bridging this gap requires identifying
% which cues distinguish competing interpretations and connecting them to
% the question being answered.

\begin{figure*}[t]
    \centering
    \includegraphics[width=\textwidth]{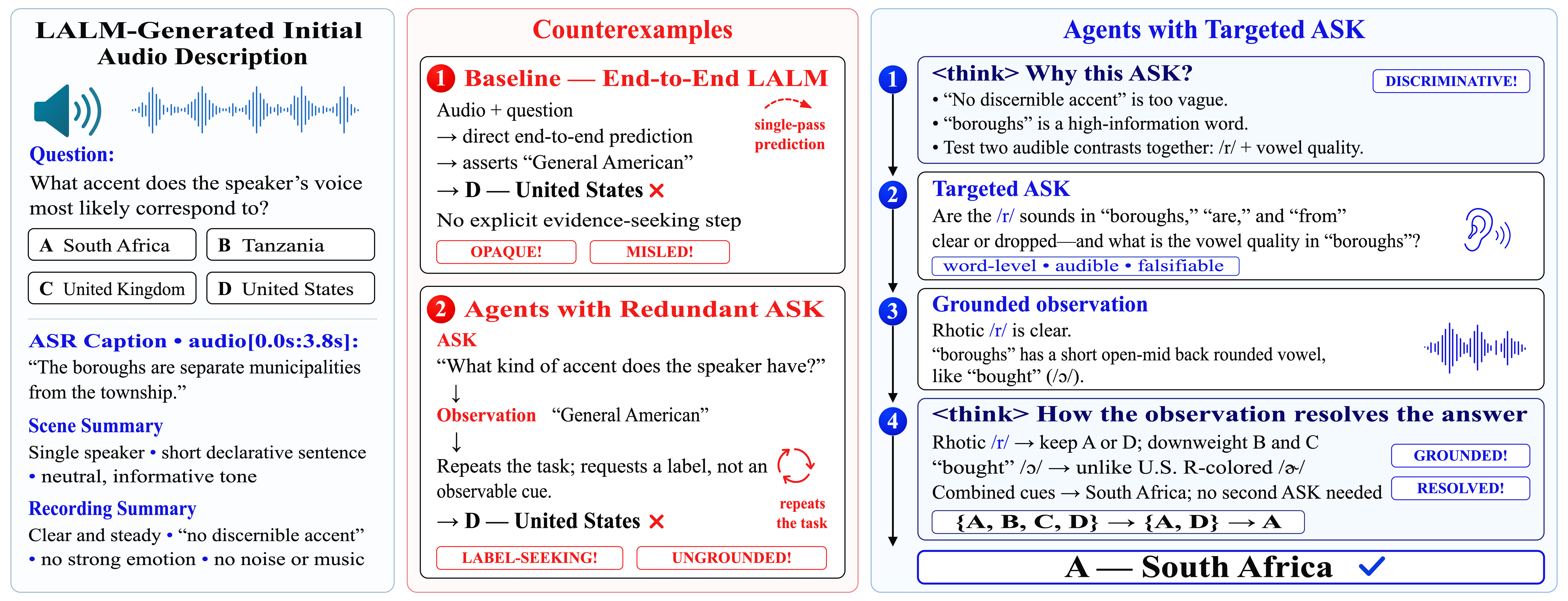}
    \caption{\textbf{An accent-identification interaction.}
    An example showing the initial audio description, an
    end-to-end prediction, a label-level ASK and its observation, and a
    targeted ASK followed by its observation and candidate-set update.}
    \label{fig:evidence-seeking-example}
\end{figure*}

Additional reasoning or interaction does not automatically bridge this
gap. Reasoning may build on textual interpretations that are insufficiently
grounded in acoustic properties~\citep{tian2025step}, while intermediate
stages of evidence collection can degrade performance relative to direct
listening~\citep{wang2026audio}.
These limitations motivate interactions that seek acoustic cues relevant
to unresolved distinctions between competing hypotheses.
For example, Figure~\ref{fig:evidence-seeking-example} presents an
accent-identification case from MMSU~\citep{wang2026mmsu}.
Asking ``Which accent?'' requests another task-level prediction, whereas
asking about /r/ realization or a word's vowel quality seeks information
about phonetic cues that can help distinguish candidate
accents~\citep{hartmann2010rhoticity,szalay2022knowledge}.
This illustrates the principle of \emph{evidence before commitment}: the agent should seek acoustic evidence that distinguishes competing hypotheses before committing to an answer.

However, a targeted query does not guarantee a justified revision of
candidate judgments.
Audio hallucination studies report confusion between semantic and
acoustic attributes, as well as errors in event identification and
temporal relations~\citep{cheng2026aha,chen2026aha}.
Misleading observations may lead an agent to exclude the correct
candidate, reducing apparent uncertainty on an unsupported premise.
This poses a challenge for policy learning: final-answer correctness
alone provides limited guidance on these revisions.
An incorrect answer does not reveal whether the agent remained uncertain
or excluded the correct candidate because of a misleading observation.
Conversely, studies of mathematical reasoning show that correct answers
can coexist with erroneous intermediate
steps~\citep{zheng2025processbench,zhang2025lessons}.
This motivates complementing outcome feedback with information about
intermediate progress~\citep{setlur2025rewarding} and, in our setting,
how acquired observations change the agent's candidate judgments.

% However, a targeted query does not guaranty a justified answer revision.
% Audio hallucination studies document the confusion between semantic and
% acoustic attributes, as well as incorrect event identities and temporal
% relations~\citep{cheng2026aha,chen2026aha}.
% Such observations may lead an agent to narrow its candidates on an
% unsupported premise, apparently resolving uncertainty while moving the
% decision in the wrong direction. Terminal correctness alone offers limited
% guidance on these revisions: studies of mathematical reasoning show
% that correct answers can coexist with erroneous intermediate
% steps~\citep{zheng2025processbench,zhang2025lessons}.
% In audio interaction, identical failure feedback can likewise cover either
% unresolved ambiguity or an observation-driven elimination of the correct
% candidate. This motivates complementing outcome feedback with information
% about intermediate progress~\citep{setlur2025rewarding} and, in our setting,
% about how acquired observations change the agent's candidate judgments.

Taken together, these limitations motivate learning \emph{evidence
acquisition and candidate revision as a coupled process}: unresolved
distinctions between hypotheses should guide subsequent queries, and
observation-induced changes in those hypotheses should inform training credit.
Building on the interleaving of reasoning and action~\citep{yao2022react},
we introduce \textbf{\texttt{Think2Listen}}, an evidence-centric learning
framework for agentic audio understanding guided by \emph{evidence before
commitment}. Practically, its \emph{Adaptive Audio Evidence Loop} allows an
actor to choose between issuing a cue-specific query and submitting an answer.
The LALM re-examines the recording for each query, while candidate revision
and the final decision remain with the actor. To guide policy learning, we introduce \emph{Transition-as-Credit},
inspired by feedback on intermediate progress~\citep{setlur2025rewarding}.
It uses changes in candidate judgments to complement answer feedback,
helping the agent learn from both its final answer and how it uses audio evidence. Our contributions can be summarized as follows:
\begin{list}{}{%
    \setlength{\leftmargin}{1.35em}%
    \setlength{\labelwidth}{0.9em}%
    \setlength{\labelsep}{0.45em}%
    \setlength{\itemindent}{0pt}%
    \setlength{\listparindent}{0pt}%
    \setlength{\topsep}{0.3em}%
    \setlength{\itemsep}{0.3em}%
    \setlength{\parsep}{0pt}%
    \setlength{\partopsep}{0pt}%
}
    \item[\ding{182}] \textbf{\textit{Evidence-Seeking Interaction.}}
    We develop an \emph{Adaptive Audio Evidence Loop} in which an actor
    formulates free-form queries to a LALM, relates the returned acoustic
    observations to competing hypotheses, and decides whether to seek further
    evidence or commit to an answer.

    \item[\ding{183}] \textbf{\textit{Transition-Aware Policy Learning.}}
    We introduce \emph{Transition-as-Credit}, which reconstructs
    observation-induced candidate-state changes and uses them for
    group-adaptive calibration of trajectory rewards. This complements
    terminal correctness with information about intermediate candidate
    revisions, distinguishing unresolved ambiguity from misleading commitment.
    Together, these components enable the agent to learn \emph{what to ask},
    \emph{how to use evidence}, and \emph{when to stop}.

    \item[\ding{184}] \textbf{\textit{Empirical Validation.}}
    Extensive evaluations on two benchmarks spanning seven audio categories
    demonstrate that \textbf{\texttt{Think2Listen}} achieves
    \textbf{(I) enhanced multi-step audio reasoning}, with gains in
    overall accuracy of 5.08\%--17.67\% over competing audio agents; and
    \textbf{(II) improved audio understanding and reasoning} across speech,
    sound, and music, boosting overall accuracy by
    2.29\%--13.10\% over the corresponding baselines.
\end{list}

\section{Related Works}
% \noindent \textbf{Large Audio-Language Models.} Large audio-language models connect acoustic perception with language modeling, supporting audio understanding, conversation, and reasoning. Their development follows several overlapping directions: (1) \emph{audio understanding and conversation}, which extends language interfaces to diverse auditory tasks, as in LTU~\citep{gong2024listen}, SALMONN~\citep{tang2024salmonn}, Qwen2-Audio~\citep{chu2024qwen2}, Audio Flamingo 2~\citep{ghosh2025audio}, MiDashengLM~\citep{dinkel2025midashenglm},  Kimi-Audio~\citep{ding2025kimi}, Step-Audio 2~\citep{wu2025step}, Audio Flamingo 3~\citep{ghosh2026audio} and SALMONN-2~\citep{yang2026salmonn}; (2) \emph{omni-modal audio understanding}, which incorporates audio alongside text, images, and video, as in GPT-4o Audio~\citep{hurst2024gpt}, Gemini 2.0 Flash~\citep{google2025gemini20flash}, Gemini 2.5 Pro~\citep{comanici2025gemini}, Qwen2.5-Omni~\citep{xu2025qwen25omnitechnicalreport}, and Qwen3-Omni~\citep{xu2025qwen3}; and (3) \emph{audio reasoning}, which strengthens inference over acoustic content through reasoning-oriented data, prompting, or post-training, as explored in GAMA~\citep{ghosh2024gama}, Audio-CoT~\citep{ma2025audio}, Audio-Reasoner~\citep{zhifei2025audio}, Step-Audio-R1~\citep{tian2025step}, Audio-Cogito~\citep{li2026audio}, Audio-DeepThinker~\citep{he2026audio}, and Audio Flamingo Next~\citep{ghosh2026audio}.

\noindent\textbf{Large Audio-Language Models.}
Large audio-language models connect acoustic perception with language modeling,
supporting audio understanding, conversation, and reasoning.
Their development follows several overlapping directions:
(1) \emph{audio understanding and conversation}, which extends language
interfaces to diverse auditory tasks, as in
GAMA~\citep{ghosh2024gama},
Qwen2-Audio~\citep{chu2024qwen2audiotechnicalreport},
DeSTA2.5-Audio~\citep{lu2026desta2},
VITA-Audio~\citep{long2025vita},
and SALMONN-2~\citep{yang2026salmonn};
(2) \emph{omni-modal audio understanding}, which incorporates audio alongside
text, images, and video, as in
Mini-Omni2~\citep{xie2024mini},
Gemini 2.0 Flash~\citep{google2025gemini20flash},
Phi-4-Multimodal~\citep{abdin2024phi},
AR-Omni~\citep{cheng2026ar},
and Qwen3-Omni~\citep{xu2025qwen3};
and (3) \emph{audio reasoning}, which strengthens inference over acoustic content through reasoning-oriented data, prompting, or post-training, as explored in Audio-CoT~\citep{ma2025audio}, Audio-Reasoner~\citep{zhifei2025audio}, Step-Audio-R1~\citep{tian2025step}, Audio-Cogito~\citep{li2026audio}, and Audio-DeepThinker~\citep{he2026audio}.

% \noindent \textbf{Large Audio-Language Models.} Large audio-language models connect acoustic perception with language modeling, supporting audio understanding, conversation, and reasoning. Their development follows several overlapping directions: (1) \emph{audio understanding and conversation}, which extends language interfaces to diverse auditory tasks, as in SALMONN~\citep{tang2024salmonn}, MiDashengLM~\citep{dinkel2025midashenglm},  Kimi-Audio~\citep{ding2025kimi}, Step-Audio 2~\citep{wu2025step}, Audio Flamingo 3~\citep{ghosh2026audio} and SALMONN-2~\citep{yang2026salmonn}; (2) \emph{omni-modal audio understanding}, which incorporates audio alongside text, images, and video, as in GPT-4o~\citep{hurst2024gpt}, Gemini 2.0 Flash~\citep{google2025gemini20flash}, Gemini 2.5 Pro~\citep{comanici2025gemini}, Qwen2.5-Omni~\citep{xu2025qwen25omnitechnicalreport}, and Qwen3-Omni~\citep{xu2025qwen3}; and (3) \emph{audio reasoning}, which strengthens inference over acoustic content through reasoning-oriented data, prompting, or post-training, as explored in Audio-CoT~\citep{ma2025audio}, Audio-Reasoner~\citep{zhifei2025audio}, Step-Audio-R1~\citep{tian2025step}, Audio-Cogito~\citep{li2026audio}, and Audio-DeepThinker~\citep{he2026audio}.

\noindent \textbf{Agentic Audio Understanding.} Building on these audio capabilities, agentic systems use external tools and iterative interaction to acquire additional information during reasoning. Existing approaches differ in how they organize this interaction: (1) \emph{tool-based reasoning} coordinates specialized audio capabilities, as in AudioGPT~\citep{huang2024audiogpt}, AudioToolAgent~\citep{wijngaard2025audiotoolagent}, and Thinking-with-Sound~\citep{xiong2025thinking}; (2) \emph{multi-agent evidence refinement} progressively enriches audio-derived textual evidence through collaboration, as in AudioGenie-Reasoner~\citep{rong2026audiogenie} and AsymAudio~\citep{rong26b_interspeech}; and (3) \emph{learned tool selection and routing} optimizes when and how external tools are invoked, as in AuTAgent~\citep{tong2026autagent} and AudioRouter~\citep{chen2026audiorouter}. In the related task of audio captioning, ACE-Cap~\citep{ma2026ace} learns active evidence acquisition for caption generation and evaluates caption utility on downstream audio QA. 

% \noindent \textbf{Process Supervision and Credit Assignment.} Multi-step reasoning and interaction also motivate feedback that reflects the contribution of intermediate decisions. Process supervision addresses this need through (1) \emph{reasoning-step supervision}, which evaluates intermediate inference steps, as in PRM800K~\citep{lightman2024let} and Math-Shepherd~\citep{wang2024math}; and (2) \emph{agent process rewards}, which assess actions in relation to task completion, as in StepAgent~\citep{deng2024novice}, AgentPRM~\citep{xi2026agentprm}, and iStar~\citep{liu2026agentic}. 

% Space between a two-column top float (including its caption) and body text.
% This does not alter the image-to-caption spacing.
\setlength{\dbltextfloatsep}{8pt plus 2pt minus 2pt}
\begin{figure*}[t]
    \centering
    \begin{minipage}[t]{0.492\textwidth}
        \vspace{0pt}
        \centering
        \includegraphics[width=\linewidth]{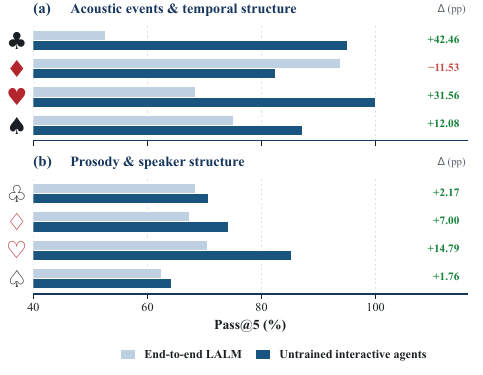}
    \end{minipage}
    \hfill
    \begin{minipage}[t]{0.492\textwidth}
        \vspace{0pt}
        \centering
        \includegraphics[width=\linewidth]{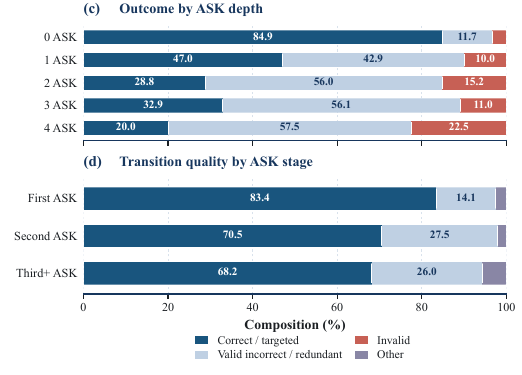}
    \end{minipage}
    \caption{\textbf{Interaction before policy training.} (a--b) Pass@5 of the
    end-to-end LALM and untrained interactive agents. Filled
    $\clubsuit/\diamondsuit/\heartsuit/\spadesuit$ denote synthetic speech,
    non-verbal sound, duration, and dialogue turns; outlined suits denote
    stress, pitch, vocal range, and speaker counting in the same order. (c--d)
    Outcomes and candidate transitions by ASK depth and stage.}
    \label{fig:interaction-diagnostics}
\end{figure*}

\section{Background \& Motivation}
\label{sec:why-agentic-audio-interaction-must-be-learned}

We contrast end-to-end prediction with interactive evidence acquisition and
examine how untrained agents use this additional audio access.

\noindent\textbf{\textit{Prediction paradigms (End-to-end and interactive inference).}}
\textit{In the end-to-end setting, a LALM predicts an answer
from the recording, task instruction, and answer
candidates}~\citep{xu2025qwen25omnitechnicalreport}.
This protocol accommodates intermediate reasoning~\citep{ghosh2026audio}
while excluding explicit queries for additional acoustic evidence.
Interactive inference extends the protocol through query-conditioned
evidence acquisition prior to the final decision.

\noindent\textbf{\textit{Empirical diagnostics (Interaction before policy training).}}
We evaluate untrained interaction along two dimensions: performance
relative to direct prediction and the outcomes of continued querying.
Figure~\ref{fig:interaction-diagnostics} summarizes the resulting patterns.

% Keep the two diagnostic points compact; spacing is local to this group.
\begingroup
\setlength{\parskip}{0.3em}
\noindent\textbullet\enspace\textbf{\textit{Task-dependent interaction gains.}}
Figure~\ref{fig:interaction-diagnostics}(a--b) compares the end-to-end LALM
with untrained interactive agents on selected MMSU tasks. Their relative
strengths vary, motivating policies that learn \emph{when to query} and
\emph{which acoustic distinctions to investigate}.

\noindent\textbullet\enspace\textbf{\textit{Continued querying and unresolved decisions.}}
Figure~\ref{fig:interaction-diagnostics}(c--d) shows that continued querying
can coexist with redundant requests and unsuccessful completion.
This does not establish that extra queries cause errors, but highlights
the need to learn \emph{how to use observations} and \emph{when to stop}.
\par
\endgroup

% Optional preamble package for the rounded, shaded version:
%   \usepackage{tcolorbox}
% Without tcolorbox, the same content uses a standard LaTeX frame.
% All definitions and frame dimensions below are local to this box.
\par\medskip
\begingroup
\long\def\AudioMotivationText{%
    \textbf{\textit{Design motivation.}}
    Reliable audio interaction requires learning \emph{what to ask},
    \emph{how to revise answers}, and \emph{when to stop}.
    Training should use changes in candidate judgments to distinguish
    useful evidence seeking from redundant or misleading interaction.}
\ifdefined\tcolorbox
    \begin{tcolorbox}[
        colback=black!3, colframe=black!60,
        boxrule=0.6pt, arc=2mm,
        left=8pt, right=8pt, top=6pt, bottom=6pt,
        before skip=0pt, after skip=0pt]
        \AudioMotivationText
    \end{tcolorbox}
\else
    \setlength{\fboxsep}{8pt}%
    \setlength{\fboxrule}{0.6pt}%
    \noindent\fbox{%
        \begin{minipage}{\dimexpr\linewidth-2\fboxsep-2\fboxrule\relax}
            \AudioMotivationText
        \end{minipage}}\par
\fi
\endgroup
\par\medskip

\section{Learning Evidence-Centric Audio Interaction}
\label{sec:learning-evidence-centric-interaction}
\begin{figure*}[t]
    \centering
    \includegraphics[width=\textwidth]{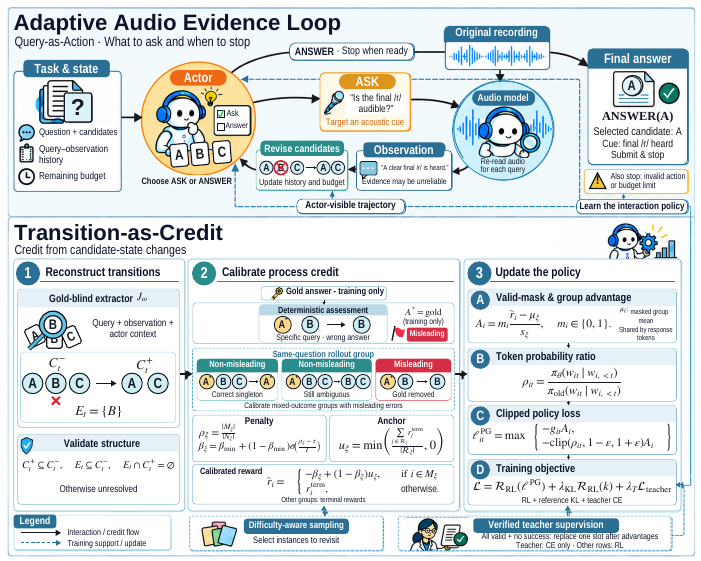}
    \vspace{-0.7em}
    \caption{\textbf{Evidence-centric audio interaction and policy learning.}
Top: the actor requests audio evidence or submits an answer. Bottom: transition extraction and gold-aware reward calibration guide policy updates. Dashed paths indicate difficulty-aware sampling and teacher supervision.}
    \label{fig:method-overview}
\end{figure*}

Section~\ref{sec:why-agentic-audio-interaction-must-be-learned} shows that
interaction can expose acoustic evidence but remains unreliable without policy
training. We therefore introduce an evidence-centric framework with two
components: the \emph{Adaptive Audio Evidence Loop} for inference and
\emph{Transition-as-Credit} for training.
Figure~\ref{fig:method-overview} summarizes the framework; the following
sections describe the inference-time evidence loop, candidate-state
reconstruction, group-adaptive credit assignment, and policy update.

\subsection{Adaptive Audio Evidence Loop}
\label{sec:adaptive-audio-evidence-loop}

The loop implements \emph{evidence before commitment} through
interaction between an actor and a LALM. The actor uses textual
evidence to decide what to ask and when to answer, while the LALM
accesses the recording to provide query-specific observations.

Given an instance $\xi=(x,q,\mathcal C)$, where $x$ is the recording,
$q$ is the task instruction, and $\mathcal C$ is the set of answer
candidates, the actor first receives an initial audio description
$d_0$ from the LALM. This description supplies the actor's initial
acoustic evidence. At turn $t$, its state is
\begin{equation}
    s_t=(q,\mathcal C,d_0,\mathrm H_t,b_t),
    \label{eq:actor-state}
\end{equation}
where $\mathrm H_t$ contains the preceding query--observation pairs
and $b_t$ is the remaining query budget. Initially,
$\mathrm H_0=\varnothing$ and $b_0=\mathrm K$, with $\mathrm K$
denoting the maximum number of additional audio queries.

The actor policy $\pi_\theta$ samples an action:
\begin{equation}
    a_t\sim\pi_\theta(\cdot\mid s_t),
    \qquad
    a_t\in\{\mathrm{ASK}(z_t),\ \mathrm{ANSWER}(c_t)\}.
    \label{eq:public-action}
\end{equation}
Here $z_t$ is a free-form audio query and $c_t\in\mathcal C$ is the
submitted answer. The query content itself is therefore a policy
action, intended to resolve an acoustic distinction between
candidates, such as a word's pronunciation or the order of two events.

For $\mathrm{ASK}(z_t)$, the audio model $p_\phi$ processes the
recording conditioned on the query and returns a textual observation
$o_t\sim p_\phi(\cdot\mid x,z_t)$.
Different queries can reveal different aspects of the same recording,
allowing the actor to seek information missing from its current context.

After each query, the history and budget are updated:
\begin{equation}
    \mathrm H_{t+1}=\mathrm H_t\oplus(z_t,o_t),
    \qquad b_{t+1}=b_t-1,
    \label{eq:state-revision}
\end{equation}
where $\oplus$ appends the query--observation pair.
Using the updated context, the actor reassesses which candidates
remain plausible and whether further evidence is needed, adapts
its next query, and makes a stopping decision accordingly.
An action $\mathrm{ANSWER}(c_t)$  submits the answer and ends the
interaction.
The next subsection describes how candidate
revisions are reconstructed for training.

\subsection{Candidate-State Transition Evaluation}
\label{sec:candidate-state-transition}

Terminal correctness alone does not establish whether an individual interaction contributed useful evidence to the reasoning process. Two trajectories may arrive at the same final answer, even though one eliminates a plausible distractor while the other merely reiterates information already available. We therefore assess the evidential contribution of interactions in terms of the resulting changes in candidate state, rather than the surface form of their queries or observations.

For each $\mathrm{ASK}$ turn selected for evaluation, a semantic extractor
$\mathrm{J}_\omega$, parameterized by $\omega$, reconstructs how the returned
observation changes the actor's candidate state:
\begin{equation}
    (v_t,\mathcal C_t^{-},\mathcal C_t^{+},\mathcal E_t)
    =\mathrm{J}_\omega(q,\mathcal C,d_0,\mathrm{H}_t,\eta_t,z_t,o_t).
    \label{eq:candidate-reconstruction}
\end{equation}
Here $q$ and $\mathcal C$ are the task instruction and answer candidates,
$d_0$ is the initial audio description, $\mathrm{H}_t$ contains the preceding
query--observation history, and $\eta_t$ is the actor reasoning associated
with the current query $z_t$. The output $v_t\in\{0,1\}$ marks whether the
query is specific and decision-relevant; $\mathcal C_t^{-}$ and
$\mathcal C_t^{+}$ are the plausible candidate sets before and after the
observation; and $\mathcal E_t$ is the set of eliminated candidates based on the returned observation.

The structured output is accepted only when its labels belong to
$\mathcal C$ and satisfy
$\mathcal C_t^{+}\subseteq\mathcal C_t^{-}$,
$\mathcal E_t\subseteq\mathcal C_t^{-}$, and
$\mathcal E_t\cap\mathcal C_t^{+}=\varnothing$; otherwise the extraction is
recorded as unresolved. Importantly, $\mathrm{J}_\omega$ receives only information
available along the trajectory and never observes the gold answer
$y^*\in\mathcal C$. It therefore performs candidate-state reconstruction rather than
reward assignment. Section~\ref{sec:group-adaptive-process-credit-calibration}
describes how these structured transitions are subsequently converted into
group-relative training credit.
\revised{This reconstruction assesses the effect of textual feedback on
candidate judgments; it does not verify the feedback against
the waveform. Answers are used afterward to flag erroneous
candidate elimination.}

\subsection{Group-Adaptive Process Credit Calibration}
\label{sec:group-adaptive-process-credit-calibration}

We deterministically flag a trajectory as \emph{misleading} if its
final answer is incorrect and at least one evaluated query with
$v_t=1$ yields an observation that contracts a two-candidate set
containing the gold answer to an incorrect singleton.
\revised{A useful query can receive inaccurate feedback, so assigning the
full failure penalty to the trajectory may suppress useful evidence seeking.
Penalty attenuation accounts for this uncertainty in attributing failure.
The actor may also misuse feedback, or reconstruction may be inaccurate;
we therefore retain a negative calibrated reward rather than treating a
flagged transition as evidence of a successful interaction.}

For each instance $\xi$, we sample $\mathrm G$ on-policy trajectories
and denote the misleading subset by $\mathcal M_\xi$. Calibration is
activated only when the group contains both correct and incorrect
trajectories and $\mathcal M_\xi\neq\varnothing$. Otherwise, terminal
rewards remain unchanged.

Within an activated group, let $\rho_\xi$ be the fraction of incorrect
trajectories flagged as misleading. This fraction determines the weight
assigned to the incorrect-answer reward of $-1$:
\begin{equation}
\beta_{\xi} =
\beta_{\min} + (1-\beta_{\min})
\sigma\bigl((\rho_{\xi}-\tau)/\mathrm{T}\bigr).
\label{eq:adaptive-penalty}
\end{equation}
Here $\sigma$ is the logistic sigmoid, $\beta_{\min}\in[0,1]$ is the
minimum weight, and $\tau$ and $\mathrm T>0$ control the midpoint and
smoothness of its dependence on $\rho_\xi$.

We define a group reference $u_\xi$ as the mean terminal reward of
the active, non-misleading trajectories, capped at zero. The calibrated
reward for trajectory $i$ is
\begin{equation}
\widetilde{r}_i =
\begin{cases}
-\beta_\xi+(1-\beta_\xi)u_\xi,
& i\in\mathcal{M}_\xi,\\
r_i^{\mathrm{term}},
& \text{otherwise}.
\end{cases}
\label{eq:adaptive-negative-reward}
\end{equation}
Here $r_i^{\mathrm{term}}$ is the original terminal reward. Misleading
trajectories receive a reward interpolated between $u_\xi$ and $-1$;
as misleading errors become more prevalent, $\beta_\xi$ increases and
the calibrated reward moves closer to $-1$. Other trajectories retain
their terminal rewards. Finally, we standardize the calibrated rewards to obtain advantages
$\mathrm A_i$ for policy learning.

\subsection{Learning the Interaction Policy}
\label{sec:learning-interaction-policy}

\noindent\textbf{Advantage-weighted policy update.}
The calibrated advantage $\mathrm A_i$ from
Section~\ref{sec:group-adaptive-process-credit-calibration} weights all
actor-generated response tokens in trajectory $i$. For token $w_{it}$
with full prefix $w_{i,<t}$, the current policy $\pi_\theta$ and rollout
policy $\pi_{\mathrm{old}}$ define the probability ratio
$\varrho_{it}=\pi_\theta(w_{it}\mid w_{i,<t})/
\pi_{\mathrm{old}}(w_{it}\mid w_{i,<t})$.
The clipped token loss is
\begin{equation}
    \ell_{it}^{\mathrm{PG}}
    =
    \max(
    -\varrho_{it}\mathrm{A}_i,
    -\mathrm{clip}(\varrho_{it},1-\epsilon,1+\epsilon)\mathrm{A}_i
    ).
    \label{eq:clipped-policy-loss}
\end{equation}
Here $\epsilon>0$ is the clipping radius.

\noindent\textbf{Verified teacher supervision.}
We supplement all-incorrect on-policy groups with available
answer-verified teacher demonstrations, following
ReLIFT~\citep{ma2026learning}. These demonstrations provide additional
reasoning examples for questions that remain unsolved in the current
rollouts. We expand demonstration coverage using targeted audio
queries and their observations as hints; generation and supervision
details are provided in Appendix~\ref{app:teacher-supervision}.

\noindent\textbf{Joint training objective.}
We combine the policy loss, reference-policy KL regularization, and
teacher supervision:
\begin{equation}
    \mathcal L
    =
    \mathcal R_{\mathrm{RL}}(\ell^{\mathrm{PG}})
    +\lambda_{\mathrm{KL}}\mathcal R_{\mathrm{RL}}(k)
    +\lambda_{\mathrm{T}}\mathcal L_{\mathrm{teacher}}.
    \label{eq:interleaved-objective}
\end{equation}
Here $\ell^{\mathrm{PG}}$ collects the token losses,
$\mathcal R_{\mathrm{RL}}$ aggregates over non-teacher on-policy
trajectories, $k$ is the reference-policy KL term, and
$\mathcal L_{\mathrm{teacher}}$ is the masked teacher cross-entropy.
The coefficients $\lambda_{\mathrm{KL}}$ and $\lambda_{\mathrm T}$
weight the KL and teacher losses, respectively. Separately, difficulty-aware sampling controls which instances are
revisited without changing rewards or this objective, as detailed in Appendix~\ref{app:difficulty-schedule}.

\vspace{-0.6em}
\section{Experiment}
\label{sec:experiment}
\vspace{-0.4em}

We organize our experimental evaluation around four research questions: \textbf{RQ1:} How does Think2Listen perform against end-to-end LALMs and existing audio agents? \textbf{RQ2:} How does policy training shape when and how agents seek additional audio evidence? \textbf{RQ3:} Does learning what to ask help agents acquire useful evidence for better audio reasoning? \textbf{RQ4:} Does transition-level credit assignment improve reliability in agentic audio reasoning?

% \begin{list}{\textbullet}{%
%     \setlength{\labelwidth}{0.65em}%
%     \setlength{\labelsep}{0.45em}%
%     \setlength{\leftmargin}{\dimexpr\labelwidth+\labelsep\relax}%
%     \setlength{\itemsep}{0.35em}%
%     \setlength{\parsep}{0pt}%
%     \setlength{\topsep}{0.25em}%
% }
%     \item \textbf{RQ1:} How does Think2Listen perform against end-to-end LALMs and existing audio agents?

%     \item \textbf{RQ2:} How does policy training shape when and how agents seek additional audio evidence?

%     \item \textbf{RQ3:} Does learning what to ask help agents acquire useful evidence for better audio reasoning?

%     \item \textbf{RQ4:} Does transition-level credit assignment improve reliability in agentic audio reasoning?
% \end{list}

\subsection{Experimental Setup}
\label{sec:experimental-setup}

\noindent\textbf{Datasets and Benchmarks.}
We evaluate Think2Listen on audio understanding and reasoning across three
domains: \emph{sound}, \emph{music}, and \emph{speech}. For \textit{evaluation},
we use two complementary benchmarks: (1) MMAU-test-mini, the
test-mini subset of MMAU~\citep{sakshi2024mmaumassivemultitaskaudio}, to assess
understanding and reasoning within each audio domain; and
(2) MMAR~\citep{ma2026mmar},
to evaluate multi-step reasoning over sound, music, speech, and their mixtures.
For \textit{policy training}, we construct a dataset of 1,770 examples
drawn from MMAU-Pro~\citep{kumar2025mmau}, which covers reasoning
across audio domains; MuChoMusic~\citep{weck2024muchomusic}, which
focuses on music understanding; and MMSU~\citep{wang2026mmsu}, which
targets spoken language understanding and reasoning. We select examples to
provide balanced coverage of sound, music, and speech.

% Local paragraph spacing keeps the results-page text compact.
\begingroup
\setlength{\parskip}{2pt}
\noindent\textbf{Method Framework.}
As shown in Figure~\ref{fig:method-overview},
\texttt{Think2Listen} assigns roles to three models.
The \textit{actor policy} ($\pi_\theta$), initialized with
Qwen3.5-9B~\citep{qwen3_5_2026}, uses the task, textual interaction
history, and remaining budget to emit $\mathrm{ASK}(z_t)$
or $\mathrm{ANSWER}(c_t)$.
The \textit{audio model} ($p_\phi$),
Qwen3-Omni-30B-A3B-Instruct~\citep{xu2025qwen3},
re-reads the original recording for each query and returns
a textual observation $o_t$.
For Transition-as-Credit, the \textit{semantic extractor}
($\mathrm{J}_\omega$), Qwen3.6-27B~\citep{qwen3.6-27b},
reconstructs candidate-state transitions for deterministic reward calibration.

\begin{table*}[t]
    \centering
    \caption{\textbf{Main results on audio understanding and reasoning benchmarks.}
    For each subcategory, \textbf{bold} and \underline{underlined} scores
    indicate the best and second-best results, respectively.
    In each Overall column, circled superscripts
    \protect\rankcircle{1}\,--\,\protect\rankcircle{5} indicate the five highest
    scores in descending order.}
    \vspace{0.3em}
    \label{tab:main-results}
    \renewcommand{\arraystretch}{1.12}
    {\fontsize{7.6}{9.0}\selectfont
    \setlength{\tabcolsep}{2.8pt}
    \begin{tabular*}{\textwidth}{@{\extracolsep{\fill}}l*{12}{c}@{}}
        \toprule
        \textbf{Method}
        & \multicolumn{8}{c}{\textbf{MMAR}}
        & \multicolumn{4}{c}{\textbf{MMAU test-mini}} \\
        \cmidrule(lr){2-9}\cmidrule(lr){10-13}
        & \textbf{S} & \textbf{M} & \textbf{Sp.}
        & \textbf{S--M} & \textbf{S--Sp.} & \textbf{M--Sp.}
        & \textbf{S--M--Sp.} & \textbf{Overall}
        & \textbf{Sound} & \textbf{Music} & \textbf{Speech}
        & \textbf{Overall} \\
        \midrule
        \rowcolor{resultbandlight}
        \multicolumn{13}{@{}l}{\textit{End-to-End LALMs}} \\
        Gemini 2.5 Flash
            & 57.42 & 48.04 & \rankfirst{79.23} & 45.45 & \rankfirst{75.37} & \ranksecond{74.67} & 54.55
            & \overallrank{67.07}{2}
            & 73.27 & 65.57 & \ranksecond{76.58} & \overallplain{71.80} \\
        Qwen3-Omni
            & 56.16 & 52.87 & 73.65 & \ranksecond{72.72} & 71.72 & 69.86 & 68.18
            & \overallrank{66.07}{3}
            & 78.72 & 71.00 & 74.10 & \overallrank{74.18}{5} \\
        GPT-4o
            & 53.94 & 50.97 & 70.41 & 63.64 & 72.48 & 62.20 & \rankfirst{75.00}
            & \overallplain{63.50}
            & 64.56 & 56.29 & 66.67 & \overallplain{62.50} \\
        MiMo-Audio
            & -- & -- & -- & -- & -- & -- & --
            & \overallplain{61.70}
            & \rankfirst{81.68} & \ranksecond{74.25} & 68.17 & \overallrank{74.70}{2} \\
        Gemini 2.0 Flash-Lite
            & 52.26 & 45.25 & 66.92 & \rankfirst{72.73} & 66.01 & 66.67 & 54.55
            & \overallplain{59.67}
            & 69.97 & 65.27 & 74.47 & \overallplain{69.90} \\
        EvA
            & -- & -- & -- & -- & -- & -- & --
            & \overallplain{59.30}
            & \ranksecond{80.78} & \rankfirst{74.65} & 68.47 & \overallrank{74.63}{3} \\
        Step-Audio-2-mini
            & -- & -- & -- & -- & -- & -- & --
            & \overallplain{56.80}
            & 79.30 & 68.44 & 68.16 & \overallplain{72.73} \\
        Qwen2.5-Omni
            & 58.79 & 40.78 & 59.86 & 54.55 & 61.93 & 67.07 & 58.33
            & \overallplain{56.70}
            & 78.10 & 65.90 & 70.60 & \overallplain{71.50} \\
        Kimi-Audio
            & 50.32 & 37.99 & 56.15 & 36.36 & 61.58 & 60.00 & 45.45
            & \overallplain{52.60}
            & 74.17 & 58.38 & 66.07 & \overallplain{66.20} \\
        Audio Flamingo 3
            & 52.26 & 32.40 & 49.62 & 27.27 & 54.19 & 52.00 & 31.82
            & \overallplain{47.18}
            & 74.47 & 67.37 & 61.26 & \overallplain{67.70} \\
        MiDashengLM
            & 43.87 & 40.22 & 51.15 & 18.18 & 45.81 & 57.33 & 36.36
            & \overallplain{46.30}
            & 69.67 & 58.98 & 62.16 & \overallplain{63.60} \\
        Audio Reasoner
            & 43.64 & 33.50 & 32.99 & 45.45 & 42.66 & 31.71 & 25.00
            & \overallplain{36.80}
            & 67.87 & 69.16 & 66.07 & \overallplain{67.70} \\
        SALMONN
            & 30.30 & 31.07 & 34.69 & 9.09 & 34.86 & 35.37 & 41.67
            & \overallplain{33.20}
            & 41.14 & 37.13 & 26.43 & \overallplain{34.90} \\
        Audio Flamingo 2
            & 24.85 & 17.48 & 20.75 & 18.18 & 26.61 & 23.17 & 8.33
            & \overallplain{21.90}
            & 71.47 & 70.96 & 44.74 & \overallplain{62.40} \\
        \midrule
        \rowcolor{resultbanddark}
        \multicolumn{13}{@{}l}{\textit{Audio Agents}} \\
        AudioToolAgent
            & 61.21 & \ranksecond{57.28} & 69.05 & 54.55 & 69.27 & \rankfirst{78.05} & 66.67
            & \overallrank{65.90}{4}
            & 78.38 & 69.46 & 75.08 & \overallrank{74.30}{4} \\
        AudioRouter
            & \ranksecond{61.40} & \rankfirst{66.10} & 77.20 & 57.20 & 65.10 & 68.30 & 55.50
            & \overallrank{64.40}{5}
            & -- & -- & -- & \overallplain{--} \\
        ACE-Cap
            & -- & -- & -- & -- & -- & -- & --
            & \overallplain{64.10}
            & -- & -- & -- & \overallplain{73.00} \\
        AuTAgent
            & 51.52 & 45.15 & 73.81 & -- & -- & -- & --
            & \overallplain{63.00}
            & 66.97 & 61.80 & 72.97 & \overallplain{67.20} \\
        AudioGenie-Reasoner
            & 49.68 & 43.26 & 69.23 & 45.45 & 64.53 & 65.33 & 59.09
            & \overallplain{58.85}
            & 75.08 & 66.17 & \ranksecond{76.58} & \overallplain{72.60} \\
        \midrule
        \textbf{Think2Listen}
            & \rankfirst{63.90} & 55.62 & \ranksecond{77.70} & 60.78 & \ranksecond{74.66} & 69.87 & \ranksecond{69.57}
            & \overallrank{69.25}{1}
            & 74.63 & 70.97 & \rankfirst{82.22} & \overallrank{76.00}{1} \\
        \bottomrule
    \end{tabular*}}
\end{table*}

\subsection{Overall Performance Comparison (RQ1)}
\label{sec:rq1}

Table~\ref{tab:main-results} compares \texttt{Think2Listen}
with end-to-end LALMs and audio agents on MMAR and MMAU test-mini.
We highlight two observations.

\begingroup
\setlength{\parskip}{2pt}
\noindent\textbf{\revised{Obs.~1. Think2Listen achieves the best overall
performance among the baselines.}}
On MMAR, \texttt{Think2Listen} achieves
\textbf{\revised{$69.25\pm0.28$\%}},
surpassing the strongest end-to-end baseline,
\texttt{Gemini 2.5 Flash}, at 67.07\% by \textbf{2.18 pp}.
On MMAU test-mini, it reaches
\textbf{\revised{$76.00\pm0.18$\%}} \revised{(mean $\pm$ standard deviation
across five evaluation rounds)},
exceeding the corresponding baseline, \texttt{MiMo-Audio},
at 74.70\% by \textbf{1.30 pp}.
It also outperforms the strongest audio agent,
\texttt{AudioToolAgent}, by \textbf{3.35 pp} and
\textbf{1.70 pp}, respectively, with larger overall gains on MMAR.
These gains extend across multi-step reasoning and broader audio understanding.
\revised{Appendix~\ref{app:additional-experiments} reports query controls,
stronger text-actor comparisons, and recent audio-model results.}

\noindent\textbf{Obs.~2. Category-level strengths vary across benchmarks.}
On MMAR, \texttt{Think2Listen} leads in sound,
surpassing \texttt{AudioRouter} by \textbf{2.50 pp}:
\(61.40\% \rightarrow \mathbf{63.90\%}\).
It also ranks second in speech, sound--speech,
and sound--music--speech, scoring 77.70\%, 74.66\%,
and 69.57\%, respectively.
On MMAU test-mini, its strongest advantage is in speech,
where it exceeds the best competing score shared by
\texttt{Gemini 2.5 Flash} and \texttt{AudioGenie-Reasoner}
by \textbf{5.64 pp}: \(76.58\% \rightarrow \mathbf{82.22\%}\).

\par
\endgroup

% Requires graphicx. Figure paths are relative to this file.
\subsection{Audio-Seeking Behavior and Feedback Value (RQ2)}
\label{sec:rq2}

We compare natural rollouts before and after training on MMSU
using identical initial audio descriptions.
A shared-prefix diagnostic further examines whether the acquired
feedback helps the policy revise its answer and make better decisions based on the additional acoustic evidence.

\begin{figure}[!htbp]
\centering
\includegraphics[width=\textwidth]{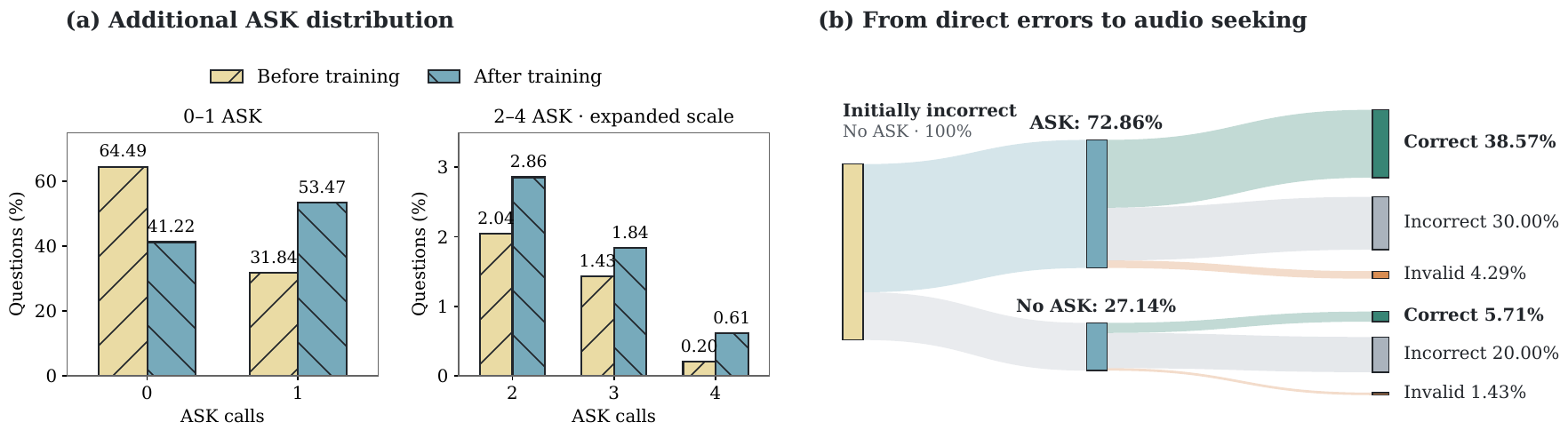}
\caption{\textbf{Natural audio seeking before and after training.}
(a) Additional \texttt{ASK} distributions;
(b) Outcomes for questions initially answered incorrectly without querying.}
\label{fig:rq2-behavior}
\end{figure}

\begingroup
\setlength{\parskip}{2pt}
\noindent\textbf{Obs.~3. Training broadens audio seeking primarily
through single-query interactions.}
The trained policy more often seeks additional audio evidence
before answering.
As shown in Figure~\ref{fig:rq2-behavior}(a),
single-query interactions become the dominant pattern,
increasing from \(31.84\% \rightarrow \mathbf{53.47\%}\),
while no-query interactions decline from 64.49\% to 41.22\%.
% \revised{Multiple-query interactions also increase, from 3.67\% to 5.31\%.
% The approximately 95\% share after training combines zero- and single-query
% interactions (41.22\% and 53.47\%); each starts with an initial audio
% description.}
This broader use of audio evidence is accompanied by higher
inclusive accuracy, \(64.08\% \rightarrow \mathbf{68.98\%}\),
with mean additional \texttt{ASK} calls remaining below one.
\revised{The budget permits four additional \texttt{ASK} calls; the separate trajectory-length penalty
is described in Appendix~\ref{app:training-evaluation-configurations}.}

\noindent\textbf{Obs.~4. Audio feedback improves decision quality
at newly queried states.}
The shift toward audio seeking also reaches questions
that were initially answered incorrectly without any querying.
After training, \textbf{72.86\%} of the questions in this cohort receive an
\texttt{ASK}, and \textbf{38.57\%} of the full cohort both
query and answer correctly, as shown in
Figure~\ref{fig:rq2-behavior}(b).
To assess the contribution of the acquired audio feedback, the shared-prefix
diagnostic holds the answering actor fixed on newly queried states.
Providing the actor with the first query--observation pair raises accuracy
from \(21.57\% \rightarrow \mathbf{47.06\%}\),
supporting the value of the acquired acoustic evidence for subsequent answer revision.
\revised{Additional cohort and paired-outcome diagnostics appear in
Appendix~\ref{app:rq2-additional-diagnostics}.}
\par
\endgroup

\subsection{Learning What to Ask for Audio Reasoning (RQ3)}
\label{sec:rq3}

We compare \textit{No ASK}, \textit{Generic query},
\textit{Untrained query}, and \textit{Learned query}
on a fixed dataset to assess whether learning what to ask
provides benefits beyond generic descriptions and untrained querying.
Detailed configurations are provided in the appendix~\ref{app:rq3-configurations}.
\revised{These comparisons evaluate complete interaction policies, including
query generation and answer reasoning, rather than isolating query learning
under a fixed answering actor. The shared-prefix diagnostic in
Section~\ref{sec:rq2} separately holds the answering actor fixed to assess
the utility of acquired feedback.}
% We compare four configurations:
% \textit{No ASK} uses only the initial audio description;
% \textit{Generic query} requests another description through a fixed,
% modality-specific prompt;
% \textit{Untrained query} and \textit{Learned query} use the base
% and trained actors, respectively, to formulate queries from the
% task and interaction history.
% All configurations are evaluated on a fixed MMSU subset where
% both untrained and trained reference policies issued a legal \texttt{ASK}.

\begingroup
\setlength{\parskip}{2pt}
\noindent\textbf{Obs.~5. Generic descriptions benefit reasoning
more than perception.}
Additional descriptions provide useful support for reasoning:
\textit{Generic query} raises Pass@1 from 42.35\% to \textbf{50.78\%}.
Perception, however, remains nearly unchanged,
\(35.43\% \rightarrow 35.69\%\).
The same contrast holds for Pass@5 in
Figure~\ref{fig:rq3-query-strategies}(a),
suggesting that additional audio access through generic descriptions
offers uneven benefits across task groups.

% Allow Figure 5 to use the next page top when it does not fit here.
\begin{figure}[!htbp]
    \centering
    \includegraphics[width=\textwidth]{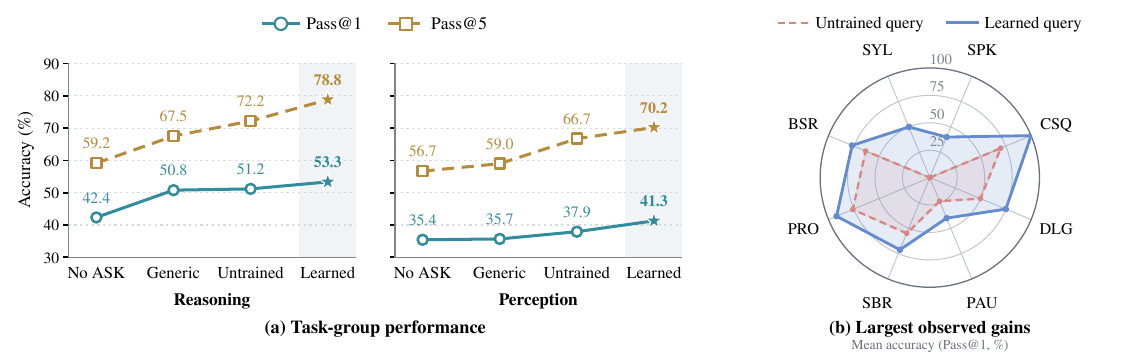}
    \caption{\textbf{Performance of audio-query strategies on MMSU.}
    (a) Pass@1 and Pass@5 by task group.
    (b) Mean Pass@1 on eight tasks, with 1--8 questions per task.
    SYL: syllables; SPK: speaker counting; CSQ: code-switching QA;
    DLG: dialogue reasoning; PAU: pauses; SBR: stress-based reasoning;
    PRO: prolonged sounds; BSR: background scenes.}
    \label{fig:rq3-query-strategies}
\end{figure}

\noindent\textbf{\revised{Obs.~6. The learned interaction policy benefits
both reasoning and perception.}}
\textit{Learned query} outperforms both generic and untrained querying,
\revised{supporting the value of jointly learning evidence acquisition
and answer revision.}
Compared with \textit{Untrained query}, Pass@1 improves
\(51.17\% \rightarrow \mathbf{53.33\%}\) for reasoning and
\(37.93\% \rightarrow \mathbf{41.29\%}\) for perception.
Relative to \textit{Generic query}, the corresponding Pass@1 gains are
\textbf{2.55 pp} and \textbf{5.60 pp}, respectively,
with the larger gain in perception highlighting the benefit of learned queries
where generic descriptions provide limited improvement.
The advantage persists under repeated sampling, with Pass@5
reaching \textbf{78.79\%} and \textbf{70.20\%}, respectively.
Figure~\ref{fig:rq3-query-strategies}(b) illustrates where
these improvements are most pronounced within the evaluated subset.
\par
\endgroup

\endgroup % Restore the template paragraph spacing.

\subsection{Training Diagnostics of Transition-as-Credit (RQ4)}
\label{sec:rq4}

\begingroup
\setlength{\parskip}{2pt}
\setlength{\intextsep}{4pt}
\begin{wraptable}{r}{0.49\textwidth}
    \vspace{-\intextsep}
    \setlength{\abovecaptionskip}{0pt}
    \setlength{\belowcaptionskip}{4pt}
    \centering

    \caption{\textbf{Validation of candidate-state reconstruction.}
    Final-answer agreement and human assessment of reconstructed transitions.}
    \label{tab:rq4-consistency}

    \renewcommand{\arraystretch}{1.18}
    {\fontsize{9}{10.5}\selectfont
    \setlength{\tabcolsep}{2pt}
    \begin{tabularx}{\linewidth}{@{}Xrr@{}}
        \toprule
        \textbf{Metric} & \textbf{TaC} & \textbf{w/o TaC} \\
        \midrule
        Parsed ASK & 20,095/20,341 & 22,247/22,352 \\
        Parse coverage & 98.79\% & 99.53\% \\
        Singleton turns & 9,273 & 9,985 \\
        Agrees with answer & 8,034 & 8,706 \\
        \rowcolor[RGB]{225,238,239}
        \textbf{Agreement} & \textbf{86.64\%} & \textbf{87.19\%} \\
        \midrule
        Human audit & 75\% & 77\% \\
        \bottomrule
    \end{tabularx}
    }

    \vspace{-4pt}
\end{wraptable}

We assess candidate-state reconstruction through \textit{answer agreement}
and \textit{human audit}, then compare Transition-as-Credit (TaC)
with a historical run without TaC on training accuracy,
misleading transitions, and query usage.
Detailed experimental settings are provided in the appendix~\ref{app:rq4-diagnostics}.
% We compare Transition-as-Credit (TaC) with a historical run
% without TaC (w/o TaC), using the same prompt schedule and
% 25,600 trajectories per run over 100 training rollouts.
% We first assess whether reconstructed candidate states provide
% a reliable basis for transition-level diagnostics.
% \textit{Answer agreement} checks whether a reconstructed
% post-\texttt{ASK} singleton $\mathcal{C}_t^{+}$ matches the
% actor's recorded final answer.
% \textit{Human audit} evaluates whether the candidate sets
% before and after an \texttt{ASK},
% $\mathcal{C}_t^{-}$ and $\mathcal{C}_t^{+}$,
% are reasonable given their corresponding interaction contexts.
% We audit 100 samples per method, accepting a sample only
% when both candidate sets are judged reasonable.
% Automatic statistics cover \texttt{ASK} turns from trajectories
% with valid actions but incorrect final answers.
% We then examine misleading candidate transitions alongside
% training accuracy and query usage.

\noindent\textbf{Obs.~7. Reconstructed candidate states are broadly
consistent with actor choices and human judgments.}
Table~\ref{tab:rq4-consistency} shows near-complete parse coverage
and comparable answer agreement across runs:
\textbf{86.64\%} for TaC and \textbf{87.19\%} for w/o TaC.
Human reviewers accept both candidate sets in 75\% and 77\%
of samples, respectively.
These results indicate that reconstructed candidate sets are
generally consistent with the actor's choices and
the available interaction evidence.

\begin{figure}[!htbp]
    \centering
    \includegraphics[width=\linewidth]{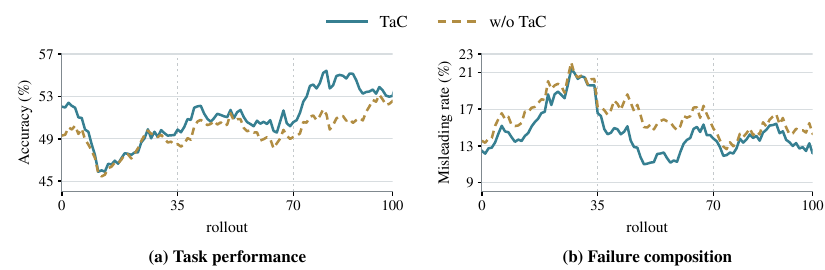}
    \caption{\textbf{Training accuracy and misleading candidate transitions.}
    (a) Accuracy over all trajectories.
    (b) Misleading trajectories as a percentage of fully terminal errors.}
    \label{fig:rq4-training}
\end{figure}

\noindent\textbf{Obs.~8. Higher training accuracy accompanies
a lower prevalence of misleading transitions.}
A misleading trajectory contains an \texttt{ASK} that contracts
a two-candidate state containing the gold answer into an incorrect
singleton and ultimately ends with an incorrect answer.
Figure~\ref{fig:rq4-training} shows that TaC generally achieves
higher final-answer accuracy and lower misleading rates than w/o TaC.
From rollouts 0--34 to 35--69,
TaC's misleading rate among fully terminal errors declines
\(16.85\% \rightarrow \mathbf{13.47\%}\),
while overall trajectory accuracy increases
\(48.52\% \rightarrow \mathbf{50.99\%}\).
The accuracy advantage also holds over the full run,
with \textbf{50.81\%} for TaC versus 50.04\% for w/o TaC,
an absolute improvement of \textbf{0.77 pp},
alongside fewer \texttt{ASK} calls on average.
% \revised{Appendices~\ref{app:verified-feedback} and
% \ref{app:extractor-robustness} further examine failures under acoustically
% verified feedback and the sensitivity of calibration to a smaller extractor.}
\par
\endgroup

% Flush outstanding result figures before the conclusion begins.
\FloatBarrier
\section{Conclusion}
In this paper, We presented \textbf{\texttt{Think2Listen}}, an evidence-centric learning
framework for agentic audio understanding guided by \emph{evidence before
commitment}. It enables an interactive reasoning paradigm in which an agent
actively seeks acoustic evidence through targeted LALM queries, revises
competing hypotheses as new evidence emerges, and decides when to commit
to an answer supported by acoustic evidence.
To learn this interaction, we utilize observation-induced candidate-state
changes to calibrate trajectory rewards, complementing terminal correctness
with feedback on intermediate evidence use.
Evaluations on two benchmarks spanning seven audio categories show gains in
multi-step audio reasoning and audio understanding across speech, sound,
and music. These results support jointly learning evidence acquisition and
candidate revision. Learning \emph{what to ask}, \emph{how to use evidence},
and \emph{when to stop} helps connect acoustic perception with
evidence-grounded decisions.

\section*{AI Use Statement}
We used AI tools to assist with writing and to polish the language of this paper.
The authors remain fully responsible for the content, experimental results,
analyses, and claims presented in this paper.

\section*{Ethics Statement}
Think2Listen uses existing audio datasets and pretrained models to study
audio understanding and reasoning. Model-generated observations and
candidate-state reconstructions may contain errors and reflect biases in
the underlying data and models. Applications involving personal audio should respect
consent and privacy and include safeguards against unsupported inferences
about individuals.

\section*{Reproducibility Statement}
Data sources and model roles are described in
Section~\ref{sec:experimental-setup}. Sources of baseline results,
teacher supervision and difficulty-aware sampling, training and decoding
settings, audio-query configurations, reconstruction-validation protocols
and metric definitions, prompts, and qualitative interaction traces are
documented in
Appendices~\ref{app:baseline-models-sources}--\ref{app:qualitative-audio-cases}.

\bibliographystyle{iclr2027_conference}
\bibliography{iclr2027_conference}

\appendix
% Keep appendix spacing natural around fixed-position tables and figures.
\raggedbottom
\section{Baseline Models and Sources of Reported Results}
\label{app:baseline-models-sources}

We select a diverse set of representative baselines for
Table~\ref{tab:main-results}, spanning end-to-end LALMs and audio agents.
The LALMs range from proprietary multimodal systems, including
Gemini 2.5 Flash~\citep{comanici2025gemini},
GPT-4o~\citep{hurst2024gpt}, and
Gemini 2.0 Flash-Lite~\citep{google2025gemini20flash},
to open models for audio understanding and reasoning, including
Qwen3-Omni~\citep{xu2025qwen3},
MiMo-Audio~\citep{zhang2025mimo},
EvA~\citep{xie2026eva},
Step-Audio-2-mini~\citep{wu2025step},
Qwen2.5-Omni~\citep{xu2025qwen25omnitechnicalreport},
Kimi-Audio~\citep{ding2025kimi},
Audio Flamingo 3~\citep{ghosh2026audio},
MiDashengLM~\citep{dinkel2025midashenglm},
Audio Reasoner~\citep{zhifei2025audio},
SALMONN~\citep{tang2024salmonn}, and
Audio Flamingo 2~\citep{ghosh2025audio}.
We further include audio agents that span tool-based interaction,
multi-agent evidence refinement, and learned tool use:
AudioToolAgent~\citep{wijngaard2025audiotoolagent},
AudioRouter~\citep{chen2026audiorouter},
ACE-Cap~\citep{ma2026ace},
AuTAgent~\citep{tong2026autagent}, and
AudioGenie-Reasoner~\citep{rong2026audiogenie}.

For MMAU, we report results on the test-mini split.

\section{Additional Methodological Details}
\label{app:methods-detail}

\subsection{Verified Teacher Supervision}
\label{app:teacher-supervision}

Following ReLIFT~\citep{ma2026learning}, we supplement unsuccessful policy groups with verified demonstrations. We expand teacher coverage
through targeted audio queries, drawing on privileged conditioning in
OPSD~\citep{zhao2026selfdistilledreasoneronpolicyselfdistillation} and audio-based supervision in
AudioRubrics~\citep{yu2026reinforcementlearningevolvingrubrics}.

We first collect answer-verified teacher trajectories, retaining one per
covered instance. For a scheduled instance $\xi$ absent from this bank,
a planner generates a hint query and obtains its audio observation:
\begin{equation}
    z_\xi^{\mathrm h}
      =\mathrm P(q,\mathcal C,y^*,\mathcal F_\xi),
    \qquad
    o_\xi^{\mathrm h}\sim
      p_\phi(\cdot\mid x,z_\xi^{\mathrm h}).
    \label{eq:teacher-hint-acquisition}
\end{equation}
Here $\mathrm P$ is the planner and $\mathcal F_\xi$ summarizes failed
attempts. Queries requesting answer labels or reproducing candidate
text are rejected. A fresh actor continues from the original run
augmented with this query--observation pair, without gold, planner
reasoning, or failed rollout history. We retain one answer-correct
continuation per newly covered instance.

During training, teacher supervision is activated by
\begin{equation}
    g_\xi=v_\xi b_\xi
      \prod_{i=1}^{\mathrm G}\mathbf{1}[r_{\xi i}=-1].
    \label{eq:teacher-activation}
\end{equation}
Here $\mathrm G$ is the rollout group size,
$r_{\xi i}\in\{-1,+1\}$ is the binary correctness reward, and
$\mathbf{1}[\cdot]$ is an indicator. The indicators $v_\xi$ and $b_\xi$
require no removed trajectories and a usable bank entry, respectively.
An activated group replaces its longest trajectory with the demonstration,
preserving the original rewards and advantages. Teacher rows contribute
the auxiliary loss
\begin{equation}
    \mathcal L_{\mathrm{teacher}}
    =-\frac{1}{B}\sum_{j\in\mathcal I_{\mathrm T}}
      \frac{\sum_t M_{jt}\log\pi_\theta(w_{jt}\mid w_{j,<t})}
           {\max(1,\sum_t M_{jt})}.
    \label{eq:teacher-masked-ce}
\end{equation}
Here $B$ is the total batch size, $\mathcal I_{\mathrm T}$ indexes teacher
rows, and $w_{jt}$ and $w_{j,<t}$ denote a target token and its full prefix.
The binary mask $M_{jt}$ selects supervised tokens. For rescued
trajectories, only actor-generated continuation actions are supervised;
hints, observations, and prompts remain context without gradients.
Policy-gradient and reference-KL losses apply only to on-policy rows
in Eq.~\eqref{eq:interleaved-objective}.

\subsection{Difficulty-Aware Sampling}
\label{app:difficulty-schedule}

Under binary rewards, uniformly correct or incorrect groups have zero
reward variance and provide no relative policy-gradient signal in
GRPO~\citep{shao2024deepseekmathpushinglimitsmathematical}.
We therefore use fixed stratified sampling to prioritize potentially
informative questions while preserving exposure to harder and unobserved
instances.

Historical performance informs difficulty-aware
selection~\citep{kong2026rethinking,zheng2026act}. We use each observed instance's
empirical success rate:
\begin{equation}
    \widehat p_\xi=\frac{c_\xi}{n_\xi}.
    \label{eq:difficulty-historical-success}
\end{equation}
Here $n_\xi>0$ counts all attempts for instance $\xi$ within a fixed
historical window, including invalid attempts, and $c_\xi$ counts those
with positive base rewards. We aggregate repeated occurrences and define
two preferential sampling pools:
\begin{equation}
    \begin{aligned}
    \mathcal D_{\mathrm H}
      &=\{\xi:n_\xi>0,\ 0<\widehat p_\xi<\alpha_{\ell}\},\\
    \mathcal D_{\mathrm M}
      &=\{\xi:n_\xi>0,\ \alpha_{\ell}\leq\widehat p_\xi
                                     \leq\alpha_u\}.
    \end{aligned}
    \label{eq:difficulty-pools}
\end{equation}
Here $\mathcal D_{\mathrm H}$ contains hard instances with nonzero success,
and $\mathcal D_{\mathrm M}$ contains medium-difficulty instances, with
thresholds $\alpha_{\ell}=0.25$ and $\alpha_u=0.75$.
We combine these groups with uniform sampling throughout the training set
$\mathcal D$:
\begin{equation}
    q_{\mathrm{sched}}(\xi)
    =\omega_{\mathrm M}\frac{\mathbf{1}[\xi\in\mathcal D_{\mathrm M}]}
                                  {|\mathcal D_{\mathrm M}|}
    +\omega_{\mathrm H}\frac{\mathbf{1}[\xi\in\mathcal D_{\mathrm H}]}
                                  {|\mathcal D_{\mathrm H}|}
    +\omega_{\mathrm U}\frac{1}{|\mathcal D|}.
    \label{eq:difficulty-source-mixture}
\end{equation}
Here $\mathbf{1}[\cdot]$ is an indicator, $|\cdot|$ denotes set size, and
$(\omega_{\mathrm M},\omega_{\mathrm H},\omega_{\mathrm U})=(0.5,0.2,0.3)$
are source weights summing to one. The uniform source retains access to
easy, unseen, and historically unsolved instances; source weights thus
differ from realized difficulty proportions.

We realize the mixture through fixed offline quotas, batch-level instance
uniqueness, and a checked repetition cap within the saved sequence.
Unlike DAPO's online filtering~\citep{yu2026dapo} and CDAS's adaptive
difficulty updates~\citep{kong2026rethinking}, our schedule is constructed
before audio rollouts and shared unchanged across configurations. This
avoids additional generation solely for batch refilling and holds instance
selection and ordering fixed, supporting controlled comparisons of method
variants and hyperparameter settings.

% Keep Figure 7 between the sampling discussion and its ablation subsection.
\begin{figure}[H]
    \centering
    \begin{minipage}[b]{0.44\linewidth}
        \centering
        \begin{minipage}[c][1.90in][c]{\linewidth}
        \centering
        \begingroup
        \small
        \setlength{\tabcolsep}{0pt}
        \renewcommand{\arraystretch}{1.35}
        \begin{tabularx}{\linewidth}{@{}Xr@{}}
            \toprule
            \textbf{Configuration} & \textbf{Acc. (\%)} \\
            \midrule
            Baseline GRPO & 66.0 \\
            Difficulty-aware sampling & 66.9 \\
            Verified teacher supervision & 68.3 \\
            Transition-as-Credit & 68.7 \\
            \midrule
            \rowcolor{resultbandlight!35}
            \textbf{Think2Listen} & \textbf{71.2} \\
            \bottomrule
        \end{tabularx}
        \endgroup
        \end{minipage}
        \par\vspace{3pt}
        {\small (a) MMSU test accuracy.\par}
    \end{minipage}\hfill
    \begin{minipage}[b]{0.54\linewidth}
        \centering
        \includegraphics[width=\linewidth]{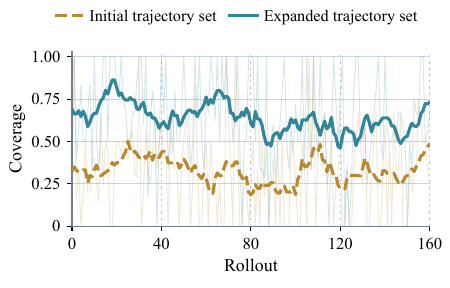}
        \par\vspace{3pt}
        {\small (b) Teacher-trajectory coverage.\par}
    \end{minipage}
    \caption{\textbf{Ablation results and teacher-trajectory coverage.}
    \revised{(a) MMSU test accuracy. The last row reports the full system's
    overall accuracy.}
    (b) Availability of verified teacher trajectories for rollout groups
    whose trajectories all receive negative terminal rewards.}
    \label{fig:appendix-ablation}
\end{figure}

\subsection{Ablation Results and Teacher-Trajectory Coverage}
\label{app:ablation-results}

Figure~\ref{fig:appendix-ablation} reports ablation results on the MMSU
test set. Transition-as-Credit reaches 68.7\% accuracy, compared with
66.0\% for baseline GRPO, while the full Think2Listen system achieves an
overall accuracy of 71.2\%. To examine verified teacher supervision
(Section~\ref{sec:learning-interaction-policy}), we compare the initial set
of answer-verified teacher trajectories with the expanded set obtained
through targeted audio queries. The expanded set maintains higher
smoothed coverage of all-negative rollout groups.

\section{Training and Evaluation Configurations}
\label{app:training-evaluation-configurations}

We initialize the actor $\pi_\theta$ and the reference policy from
Qwen3.5-9B and train on two NVIDIA B200 GPUs.
Table~\ref{tab:training-hyperparameters} summarizes the optimization,
interaction, and decoding parameters.

\begin{rebuttalrevision}
\noindent\textbf{Evaluation metric and repeated sampling.}
Our evaluations for Table~\ref{tab:main-results} estimate single-response
accuracy with the unbiased Pass@1 estimator. For question $j$, we sample
five responses and count $c_j$ correct answers, giving
$\widehat{\mathrm{Pass@1}}=\frac{1}{N}\sum_{j=1}^{N}c_j/5$ over $N$
questions. Repeated sampling estimates the performance of one sampled
response; it does not select the best answer or apply majority voting.
Think2Listen achieves $(69.25\pm0.28)$\% on MMAR and
$(76.00\pm0.18)$\% on MMAU test-mini. These values are the mean and
standard deviation of the overall accuracies across five evaluation
rounds, each using one sampled response per question. The mean is
equivalent to the Pass@1 estimator above.
Externally sourced baselines retain the reporting protocols of their
sources (Appendix~\ref{app:baseline-models-sources}); supplementary controls
with one sample per question are explicitly identified in
Appendix~\ref{app:benchmark-query-controls}.

\noindent\textbf{Note on MMSU data separation.}
We divide MMSU into disjoint training, validation, and test partitions
before constructing the mixed training set. Its 755 MMSU examples are
selected exclusively from the training partition. The analyses in
Sections~\ref{sec:rq2} and~\ref{sec:rq3} and
Appendix~\ref{app:ablation-results} draw on the same held-out test
partition of 490 questions; RQ3 uses selected questions from this partition.
The stored training and test sample lists share neither sample IDs nor
audio filenames. Validation data, rather than test results, determine
the checkpoint choices in Figure~\ref{fig:appendix-ablation}.
\end{rebuttalrevision}

% Keep the parameter table next to its introduction.
\begin{table}[H]
    \centering
    \caption{\textbf{Training and evaluation parameters.}}
    \label{tab:training-hyperparameters}
    \begingroup
    \small
    \setlength{\tabcolsep}{6pt}
    \renewcommand{\arraystretch}{1.10}
    \begin{tabularx}{\linewidth}{@{}Xl@{}}
        \toprule
        \textbf{Parameter} & \textbf{Value} \\
        \midrule
        Training epochs & 5 \\
        Prompts per rollout batch & 32 \\
        Trajectories per prompt, $\mathrm G$ & 8 \\
        Global batch size & 256 \\
        Advantage estimator & GRPO \\
        Optimizer & Adam \\
        Learning rate / schedule & $10^{-6}$ / constant \\
        Adam moments / weight decay & $(0.9,0.98)$ / 0 \\
        Clipping radius, $\epsilon$ & 0.2 \\
        KL coefficient, $\lambda_{\mathrm{KL}}$ & 0.001 \\
        KL estimator & \texttt{low\_var\_kl} \\
        Entropy coefficient & 0 \\
        Dynamic batching budget per GPU & 8,192 tokens \\
        Maximum additional $\mathrm{ASK}$ calls, $\mathrm K$ & 4 \\
        Checkpoint and evaluation interval & 20 rollout batches \\
        Actor sampling temperature during training & 1.0 \\
        Actor sampling temperature / top-$p$ at evaluation & 0.7 / 0.95 \\
        \bottomrule
    \end{tabularx}
    \endgroup
\end{table}

Inspired by the overlong reward shaping in DAPO~\citep{yu2026dapo},
we use total trajectory length to determine a length penalty:
within the 4,096--8,192-token interval, each additional 1,024 tokens
beyond 4,096 contributes a further $-0.05$ to the reward.
Trajectories exceeding 8,192 tokens are excluded from reward computation
and policy updates. At evaluation time, temperature 0.7 controls
sampling sharpness, and top-$p$ 0.95 restricts sampling to the smallest
set of most probable tokens covering at least 95\% of the probability mass.

For Transition-as-Credit, $\beta_{\min}=0.5$ sets the minimum weight
on the incorrect-answer reward of $-1$ in
Eq.~\eqref{eq:adaptive-penalty}; $\tau=0.5$ places the sigmoid midpoint
where half of the group's incorrect trajectories are misleading, and
$\mathrm T=0.1$ controls the smoothness of this transition.
The semantic extractor $\mathrm J_\omega$ with
temperature 0 and top-$p$ 0.95 for greedy decoding, with an output limit
of 8,192 tokens. For verified teacher supervision,
$\lambda_{\mathrm T}=0.1$ scales the masked teacher cross-entropy in
Eq.~\eqref{eq:interleaved-objective}. 

\section{Audio-Query Configurations for RQ3}
\label{app:rq3-configurations}

\noindent\textbf{Evaluation protocol.}
We evaluate four configurations on selected MMSU questions, with ten runs
per question per configuration, yielding 6,680 evaluation runs in total.

\noindent\textbf{Query configurations.}
\textit{No ASK} uses the base actor to answer from the initial audio
description without additional audio calls.
\textit{Generic query} selects a predefined generic query template
according to the audio type associated with each question,
as shown in Table~\ref{tab:rq3-generic-templates}.
\textit{Untrained query} allows the base actor to formulate queries
from the task and interaction history, with up to four additional audio calls.
\textit{Learned query} uses the trained actor for both query generation
and final answering, conditioned on the evolving interaction history.

\begin{table}[H]
    \centering
    \caption{\textbf{Modality-specific prompts for generic audio queries.}}
    \vspace{0.3em}
    \label{tab:rq3-generic-templates}
    \begingroup
    \small
    \setlength{\tabcolsep}{6pt}
    \renewcommand{\arraystretch}{1.12}
    \renewcommand{\tabularxcolumn}[1]{p{#1}}
    \begin{tabularx}{\linewidth}{@{}>{\raggedright\arraybackslash}p{0.14\linewidth}>{\raggedright\arraybackslash}X@{}}
        \toprule
        \textbf{Modality} & \textbf{Prompt} \\
        \midrule
        Speech & Describe the audible speech content, speaker count, and relevant background scene. State only facts that can be heard. \\
        \addlinespace[5pt]
        Music & Describe the audible musical sources and texture: instruments, voices, timbre, accompaniment, and production effects. \\
        \addlinespace[5pt]
        Sound & Identify the audible sound sources, events, and background ambience. Distinguish foreground from background when possible. \\
        \addlinespace[5pt]
        Mixed & Inventory the audible speech, music, and non-speech sound sources, separating foreground and background. \\
        \bottomrule
    \end{tabularx}
    \endgroup
\end{table}

\section{Reconstruction Validation for RQ4}
\label{app:rq4-diagnostics}

\noindent\textbf{Evaluation protocol.}
We compare Transition-as-Credit (TaC) with a historical run without TaC
(w/o TaC), using the same prompt schedule and 25,600 trajectories per run
over 100 training rollouts.
\revised{Both runs use the same setup and training settings, with TaC disabled in the historical comparison run.}
We assess candidate-state reconstruction
through answer agreement and human audit, and examine training accuracy,
misleading trajectories, and query usage.

\begin{rebuttalrevision}
\noindent\textbf{Observed calibration frequency.}
In the available training audit, 17.47\% of eight-trajectory prompt groups
contain at least one rescaled trajectory, and 5.26\% of trajectory records
are rescaled. A record is counted as rescaled when its logged calibration
scale differs from one. These are empirical audit frequencies; the
activation rule remains that a group contains both correct and incorrect
trajectories and at least one trajectory flagged as misleading
(Section~\ref{sec:group-adaptive-process-credit-calibration}).
\end{rebuttalrevision}

\noindent\textbf{Candidate-state reconstruction.}
We use the semantic extractor $\mathrm{J}_\omega$ defined in
Section~\ref{sec:candidate-state-transition} and
Eq.~\ref{eq:candidate-reconstruction}.
For each evaluated $\mathrm{ASK}$ turn, it returns
$(v_t,\mathcal C_t^{-},\mathcal C_t^{+},\mathcal E_t)$ from the task,
answer candidates, initial audio description, preceding interaction history,
actor reasoning, query, and observation.
Here $v_t$ indicates whether the query is specific and decision-relevant;
$\mathcal C_t^{-}$ and $\mathcal C_t^{+}$ are the plausible candidate sets
before and after the observation; and $\mathcal E_t$ contains the
eliminated candidates.
The extractor does not observe the gold answer $y^*$.
Following the same acceptance rule, all output candidate labels must
belong to $\mathcal C$, with
$\mathcal C_t^{+}\subseteq\mathcal C_t^{-}$,
$\mathcal E_t\subseteq\mathcal C_t^{-}$, and
$\mathcal E_t\cap\mathcal C_t^{+}=\varnothing$.
Outputs that cannot be parsed or fail these checks are recorded as unresolved.

\noindent\textbf{Answer agreement.}
Automatic reconstruction statistics cover $\mathrm{ASK}$ turns from
trajectories with valid actions but incorrect final answers.
Parse coverage is the fraction of these turns with an accepted extraction.
Among accepted extractions with $|\mathcal C_t^{+}|=1$, we compare the sole
remaining candidate with the actor's recorded final answer.
Agreement is the fraction of these singleton turns that match that answer,
with each eligible turn counted separately.
The reference is the actor's final choice, rather than $y^*$, so this check
measures consistency with the actor's decision rather than task correctness.

\noindent\textbf{Human audit.}
We manually assess 100 sampled transitions per method.
For each sample, reviewers judge whether $\mathcal C_t^{-}$ and
$\mathcal C_t^{+}$ are reasonable given the interaction context before
and after the observation, respectively.
A sample passes only when both candidate sets are judged reasonable;
the reported audit rate is the fraction of the 100 samples that pass.
This assessment complements answer agreement by checking the plausibility
of both reconstructed states, rather than only agreement with a final choice.

% BEGIN REBUTTAL ADDITIONS: all experimental supplements precede Prompts and Examples.
\begin{rebuttalrevision}
\section{Additional Evaluation and Control Experiments}
\label{app:additional-experiments}

\subsection{Query Controls on the Main Benchmarks}
\label{app:benchmark-query-controls}

We add query controls on MMAR and MMAU test-mini questions using
the Qwen3.5-9B actor and the same cached initial descriptions. Query-enabled
settings use Qwen3-Omni-30B-A3B-Instruct with up to four additional
\texttt{ASK} calls. The two trained-actor settings use the same checkpoint.
The untrained-query and trained-description-only controls use one sample
per question; learned-query results use the five-sample Pass@1 estimator
from Table~\ref{tab:main-results}. 

\begin{table}[H]
    \arrayrulecolor{black}
    \centering
    \caption{\textbf{Query controls on MMAR and MMAU test-mini.}
    Accuracy (\%); query-policy settings share cached initial descriptions.
    Sampling counts differ as specified in the evaluation protocol.}
    \label{tab:benchmark-query-controls}
    \small
    \setlength{\tabcolsep}{4pt}
    \renewcommand{\arraystretch}{1.15}
    \begin{tabular*}{\linewidth}{@{\extracolsep{\fill}}lrr@{}}
        \toprule
        \textbf{Configuration} & \textbf{MMAR} & \textbf{MMAU test-mini} \\
        \midrule
        Qwen3-Omni (direct answering) & 66.07 & 74.18 \\
        Untrained query & 60.00 & 68.82 \\
        Initial description only (trained actor) & 62.83 & 73.03 \\
        Learned query (trained actor) & 69.25 & 76.00 \\
        \bottomrule
    \end{tabular*}
\end{table}

Learned query exceeds untrained query by 9.25 and 7.18 percentage points
on MMAR and MMAU test-mini, respectively. With the trained checkpoint
fixed, allowing additional queries improves the reported accuracy by
6.42 and 2.97 points over answering from the initial description alone.
These comparisons support the value of learning the interaction policy
and acquiring additional evidence, while their sampling counts differ.
They address querying behavior on MMAR/MMAU; the TaC-specific ablation
remains the MMSU comparison in Appendix~\ref{app:ablation-results}.

\subsection{Repeated Answering and Matched Audio-Call Controls}
\label{app:budget-controls}

We also test majority voting over Qwen3-Omni's direct answers. Relative to
direct answering, voting yields two fewer correct answers on MMAR and
three fewer on MMAU test-mini, offering no accuracy gain in this experiment.

To compare targeted querying with repeated generic descriptions, we
evaluate independent audio descriptions followed by untrained-actor
answer voting on fixed subsets of 200 questions per dataset. For each
question, this control uses the same number of audio calls as its
untrained-query trajectory, including the initial description.
Table~\ref{tab:matched-audio-calls} shows that untrained querying exceeds
description-based voting by 2.00 points on MMAR and 6.00 on MMAU test-mini
under the matched audio-call budget. This matching concerns audio calls,
not total inference compute.

\begin{table}[H]
    \arrayrulecolor{black}
    \centering
    \caption{\textbf{Matched audio-call comparison.}
    Accuracy (\%) on the same fixed 200-question subset of each dataset.}
    \label{tab:matched-audio-calls}
    \small
    \setlength{\tabcolsep}{4pt}
    \renewcommand{\arraystretch}{1.15}
    \begin{tabularx}{\linewidth}{@{}Xrr@{}}
        \toprule
        \textbf{Configuration} & \textbf{MMAR} & \textbf{MMAU test-mini} \\
        \midrule
        Untrained query & 58.00 & 73.50 \\
        Independent descriptions + untrained-actor voting & 56.00 & 67.50 \\
        \bottomrule
    \end{tabularx}
\end{table}

\clearpage
\subsection{Additional Cohort and Paired-Outcome Diagnostics for RQ2}
\label{app:rq2-additional-diagnostics}

The cohort in Figure~\ref{fig:rq2-behavior}(b) comprises questions initially
answered incorrectly without querying. On this cohort, the existing
untrained-actor rerun achieves 15.71\% accuracy, compared with 44.29\%
for the trained natural policy. The rerun allows at most two repeated
label-level queries, whereas the natural trajectories allow up to four
\texttt{ASK} calls. This is therefore a descriptive comparison, rather
than a control with identical query budgets.

Across the full paired evaluation set, 5.51\% of questions change from
correct to incorrect after training, while 10.41\% change from incorrect
or invalid to correct. The net gain is 4.90 percentage points, consistent
with the overall accuracy increase from 64.08\% to 68.98\% in Obs.~3.
These paired changes complement the selected-cohort analysis. The
shared-prefix diagnostic in Obs.~4 separately fixes the answering actor
and measures the contribution of the first query--observation pair.

\subsection{Failure Recovery with Acoustically Verified Feedback}
\label{app:verified-feedback}

\noindent\textbf{Protocol.}
We re-evaluate originally incorrect answers on MMAR and MMAU test-mini to
examine the contribution of audio feedback. We regenerate both the initial
descriptions and query observations with Gemini 2.5 Pro; human listening
confirms that the replacement observations match the audio. We keep the
actor, recorded query sequence, and final-answer template fixed, and allow
no additional queries. As a control for recovery from re-answering alone,
we also ask the same actor to answer once more using the original feedback.

\noindent\textbf{Results and metric definitions.}
All rates in Table~\ref{tab:verified-feedback} use the corresponding original
failure set as the denominator. Replacing feedback recovers 38.71\% of
MMAR failures and 37.50\% of MMAU test-mini failures, whereas re-answering
with the original feedback recovers 0.00\% and 4.17\%, respectively.
We report the difference between these recovery rates as the
\emph{net recovery gain}; thus, the MMAU test-mini gain is 33.33 percentage
points, distinct from its 37.50\% total verified-feedback recovery rate.

\begin{table}[H]
    \arrayrulecolor{black}
    \centering
    \caption{\textbf{Recovery on the original failure sets.}
    Original and verified columns give recovery rates (\%); net gain is
    their difference in percentage points. Residual is the percentage still
    incorrect with verified feedback. Combined pools the failure sets.}
    \label{tab:verified-feedback}
    \small
    \setlength{\tabcolsep}{5pt}
    \renewcommand{\arraystretch}{1.15}
    \begin{tabular*}{\linewidth}{@{\extracolsep{\fill}}lrrrr@{}}
        \toprule
        \textbf{Failure set} & \textbf{Original} & \textbf{Verified}
        & \textbf{Net gain} & \textbf{Residual} \\
        \midrule
        MMAR & 0.00 & 38.71 & 38.71 & 61.29 \\
        MMAU test-mini & 4.17 & 37.50 & 33.33 & 62.50 \\
        Combined & 1.82 & 38.18 & 36.36 & 61.82 \\
        \bottomrule
    \end{tabular*}
\end{table}

The pooled net recovery gain is 36.36 percentage points of the original
failures without changing the actor or query sequence. The remaining
61.82\% persist with acoustically verified feedback, suggesting limitations
in reasoning, evidence use, or query coverage. This is a controlled
feedback-replacement diagnostic: the net gain is a difference between
recovery rates, not a per-example causal assignment of each failure to
perception or reasoning. In particular, fixing the query sequence leaves
missing evidence outside its coverage unaddressed.

\clearpage
\subsection{Comparison with Stronger Pretrained Text Actors}
\label{app:stronger-actors}

\noindent\textbf{Protocol.}
We evaluate GPT-6.1 Terra and GPT-6.1 Sol at medium, high, and xhigh
reasoning effort without task-specific training. All actors use the same
Qwen3-Omni-30B-A3B-Instruct audio model, tool interface, prompts, cached
initial descriptions, and budget of four additional audio queries.
Every result in Table~\ref{tab:stronger-actors}, including Think2Listen,
uses the same fixed subset of 100 questions from MMAR and 100 questions
from MMAU test-mini. These subset results are reported separately from
the main benchmark comparison in Table~\ref{tab:main-results}.

\begin{table}[H]
    \arrayrulecolor{black}
    \centering
    \caption{\textbf{Text-actor comparison with a shared audio interface.}
    Accuracy (\%) on fixed 100-question subsets of each benchmark.}
    \label{tab:stronger-actors}
    \small
    \setlength{\tabcolsep}{5pt}
    \renewcommand{\arraystretch}{1.15}
    \begin{tabular*}{\linewidth}{@{\extracolsep{\fill}}llrr@{}}
        \toprule
        \textbf{Text actor} & \textbf{Reasoning effort}
        & \textbf{MMAR} & \textbf{MMAU test-mini} \\
        \midrule
        Think2Listen (trained 9B) & -- & 69 & 76 \\
        GPT-6.1 Terra & medium & 69 & 76 \\
        GPT-6.1 Terra & high & 70 & 77 \\
        GPT-6.1 Terra & xhigh & 73 & 76 \\
        GPT-6.1 Sol & medium & 73 & 77 \\
        GPT-6.1 Sol & high & 75 & 74 \\
        GPT-6.1 Sol & xhigh & 76 & 79 \\
        \bottomrule
    \end{tabular*}
\end{table}

The trained 9B actor matches Terra at medium effort on both subsets and
is within one percentage point of Terra at high effort. Sol at xhigh
effort exceeds Think2Listen by 7 percentage points on MMAR and 3 on
MMAU test-mini. These results place the trained actor's performance in
context under a shared audio interface; the comparison does not equate
the actors' inference compute or establish statistical equivalence.

\subsection{Sensitivity to a Smaller Semantic Extractor}
\label{app:extractor-robustness}

\noindent\textbf{Replay protocol.}
We replace the semantic extractor with Qwen3-8B and replay the recorded
interactions to assess changes in parse coverage and calibrated
advantages. Parse coverage is compared on the same-budget subset;
advantage statistics use the full replay against the recovered 27B
reference. These are different evaluation scopes and are reported
separately in Table~\ref{tab:extractor-robustness}. This replay evaluates
the credit signal without retraining the actor with the smaller extractor.

\begin{table}[H]
    \arrayrulecolor{black}
    \centering
    \caption{\textbf{Smaller-extractor replay.}
    Advantage differences compare the 8B replay with the recovered 27B
    reference; the restricted mean uses groups with reward differences.}
    \label{tab:extractor-robustness}
    \small
    \setlength{\tabcolsep}{5pt}
    \renewcommand{\arraystretch}{1.15}
    \begin{tabularx}{\linewidth}{@{}Xr@{}}
        \toprule
        \textbf{Metric} & \textbf{Result} \\
        \midrule
        Parse coverage on same-budget subset: original $\to$ 8B
            & 98.93\% $\to$ 87.23\% \\
        Mean absolute advantage difference: overall & 0.0079 \\
        Mean absolute advantage difference: groups with reward differences
            & 0.0544 \\
        Advantage-sign changes & 0 \\
        \bottomrule
    \end{tabularx}
\end{table}

Although the smaller extractor lowers parse coverage, the evaluated
advantage directions remain unchanged. This supports stability of the
credit signal in this replay, rather than identical reconstruction or
unchanged downstream performance after retraining. Extracted structures
remain subject to the validity checks in
Section~\ref{sec:candidate-state-transition}, and eligible calibrated
rewards remain negative. The downstream TaC ablation in
Appendix~\ref{app:ablation-results} separately reports 68.7\% MMSU test
accuracy versus 66.0\% for baseline GRPO, evaluated at checkpoints
selected using validation performance before training step 160. Together with the 75\% human acceptance rate
in Table~\ref{tab:rq4-consistency}, these results document benefits at an
operating point with imperfect reconstruction.

\clearpage
\subsection{Reported Results for Recent Audio Systems}
\label{app:recent-audio-systems}

Table~\ref{tab:recent-audio-systems} supplements the main comparison with
recent audio systems. The external scores retain the protocols of their
source papers, and each dataset's source is identified separately. In
particular, the MMAU test-mini scores for Step-Audio-R1 and AF-Next-Think
come from independent evaluations in SKAD and Audio-R1, respectively.

\begin{table}[H]
    \arrayrulecolor{black}
    \centering
    \caption{\textbf{Published results for recent audio systems.}
    Accuracy (\%); source links identify the paper and table for each score.
    AF-Next denotes Audio Flamingo Next.}
    \label{tab:recent-audio-systems}
    \small
    \setlength{\tabcolsep}{4pt}
    \renewcommand{\arraystretch}{1.20}
    \begin{tabularx}{\linewidth}{@{}lrr>{\raggedright\arraybackslash}X@{}}
        \toprule
        \textbf{Model} & \textbf{MMAR} & \textbf{MMAU mini}
        & \textbf{Source} \\
        \midrule
        Think2Listen & 69.25 & 76.00 & Table~\ref{tab:main-results} \\
        Step-Audio-R1 & 69.15 & 71.10 &
        MMAR: \href{https://arxiv.org/html/2606.01802v1}{MOSS-Audio}, Table~1;
        MMAU: \href{https://arxiv.org/html/2610.28146v1}{SKAD}, Table~2 \\
        SALMONN-2 9B & 64.50 & -- &
        \href{https://arxiv.org/html/2607.17079v1}{SALMONN-2}, Table~VI \\
        SALMONN-2 30B-A3B & 67.60 & -- &
        \href{https://arxiv.org/html/2607.17079v1}{SALMONN-2}, Table~VI \\
        AF-Next Instruct & 59.70 & -- &
        \href{https://arxiv.org/html/2604.10905v1}{Audio Flamingo Next}, Table~1 \\
        AF-Next Think & 61.00 & 74.50 &
        MMAR: \href{https://arxiv.org/html/2604.10905v1}{Audio Flamingo Next}, Table~1;
        MMAU: \href{https://arxiv.org/html/2610.34609v1}{Audio-R1}, Table~1 \\
        AF-Next Captioner & 63.00 & -- &
        \href{https://arxiv.org/html/2604.10905v1}{Audio Flamingo Next}, Table~1 \\
        \bottomrule
    \end{tabularx}
\end{table}

Think2Listen's reported MMAR accuracy is numerically close to
Step-Audio-R1's 69.15\%, and its MMAU test-mini score exceeds the two
external test-mini results listed here. Differences in evaluation
protocols preclude interpreting these margins as controlled comparisons
or evidence of statistical superiority.

\end{rebuttalrevision}
% END REBUTTAL ADDITIONS

\clearpage
\section{Prompts}
\label{app:prompts}

We provide the prompts for actor action selection, initial audio description,
targeted audio observation, and candidate-state reconstruction.
Each box separates the system message from the user message or template.
Braced placeholders are populated at runtime; the original audio accompanies
the two audio-model prompts. The candidate-state extractor receives only
actor-visible information and has no access to the gold answer.

\begingroup
\raggedbottom
\definecolor{promptActor}{HTML}{2B6F94}
\definecolor{promptAudio}{HTML}{347F87}
\definecolor{promptExtractor}{HTML}{557C78}
\definecolor{promptInk}{HTML}{253238}
\definecolor{promptRule}{HTML}{CEDADC}
\lstdefinestyle{paperprompt}{%
    language={},
    basicstyle=\ttfamily\fontsize{9}{10.4}\selectfont\color{promptInk},
    columns=fullflexible, keepspaces=true,
    breaklines=true, breakatwhitespace=true,
    breakindent=0pt, breakautoindent=false,
    showstringspaces=false, showspaces=false, showtabs=false,
    upquote=true, tabsize=2,
    aboveskip=0pt, belowskip=0pt,
    xleftmargin=0pt, xrightmargin=0pt,
    numbers=none, frame=none,
}
\newcommand{\PromptGap}{\par\vspace{2pt}}
\newcommand{\PromptRole}[1]{%
    \par\noindent{\rmfamily\bfseries\fontsize{7.4}{9}\selectfont
    \color{promptAccent}#1}\par\nobreak\vspace{1pt}}
\newcommand{\PromptDivider}{%
    \par\vspace{5pt}\noindent{\color{promptRule}\rule{\linewidth}{0.35pt}}%
    \par\vspace{3pt}}
\newtcolorbox{paperpromptbox}[3]{%
    enhanced, breakable,
    colframe=#1, colback=#1!2!white, colbacktitle=#1,
    coltitle=white, coltext=promptInk,
    boxrule=0.65pt, arc=1.2pt, outer arc=1.2pt,
    boxsep=0pt, left=9pt, right=9pt, top=7pt, bottom=8pt,
    lefttitle=9pt, righttitle=9pt, toptitle=4pt, bottomtitle=4pt,
    fonttitle=\bfseries\fontsize{10}{12}\selectfont,
    title={#2},
    title after break={#2\hfill{\normalfont\rmfamily\fontsize{7.5}{9}\selectfont CONTINUED}},
    before upper={\colorlet{promptAccent}{#1}\setlength{\parskip}{0pt}},
    before skip=10pt, after skip=10pt,
    pad at break*=6pt,
    lines before break=5,
}

\begin{paperpromptbox}{promptActor}{Actor action selection}{1}
\PromptRole{SYSTEM}
% BEGIN MESSAGE 1
\begin{lstlisting}[style=paperprompt]
You are an active audio-perception agent.
You cannot hear the audio. You receive a task question, its candidate answers, an initial audio description, and any prior ASK plus Omni observation pairs.
\end{lstlisting}
\PromptGap
\begin{lstlisting}[style=paperprompt]
On every turn, output exactly one action and nothing else:
<ask>one targeted question about audible evidence</ask>
or
<answer>X</answer>
\end{lstlisting}
\PromptGap
\begin{lstlisting}[style=paperprompt]
For <answer>, replace X with exactly one listed answer label.
For <ask>, request factual evidence that can be heard in the original audio. Do not ask Omni to choose an answer, name an answer label, enumerate candidate answers, or explain which candidate is correct. Do not output reasoning, Markdown, or text outside the action tag.
\end{lstlisting}
\PromptGap
\begin{lstlisting}[style=paperprompt]
Ask only when another audible fact would materially improve the decision. When the available evidence is sufficient, output <answer> instead of another ASK.
\end{lstlisting}
% END MESSAGE 1
\PromptDivider
\PromptRole{USER TEMPLATE}
% BEGIN MESSAGE 2
\begin{lstlisting}[style=paperprompt]
[Task Question]
{question}
\end{lstlisting}
\PromptGap
\begin{lstlisting}[style=paperprompt]
[Candidate Answers]
{choices}
\end{lstlisting}
\PromptGap
\begin{lstlisting}[style=paperprompt]
[Initial Omni Description]
{initial_description}
\end{lstlisting}
\PromptGap
\begin{lstlisting}[style=paperprompt]
[Prior Qwen ASK and Omni Observation Pairs]
{interaction_history}
\end{lstlisting}
\PromptGap
\begin{lstlisting}[style=paperprompt]
Choose the next action now.
\end{lstlisting}
% END MESSAGE 2
\end{paperpromptbox}

\begin{paperpromptbox}{promptAudio}{Initial audio description}{2}
\PromptRole{SYSTEM}
% BEGIN MESSAGE 3
\begin{lstlisting}[style=paperprompt]
You are an audio analysis assistant.
Your task is to answer user questions strictly based on the factual content of the provided audio clips. Carefully listen to and analyze the audio before answering.
\end{lstlisting}
\PromptGap
\begin{lstlisting}[style=paperprompt]
Audio reference notation:
- A bare reference like <audio> refers to the full duration of that audio clip.
- A reference with [start:end] denotes a time slice of that clip (e.g. <audio[0.0s:3s]> means the segment from 0.0s to 3.0s).
- An omitted start means the beginning of the clip (e.g. <audio[:2.0s]> means from the beginning up to 2.0s).
- An omitted end means the end of the clip (e.g. <audio[1.5s:]> means from 1.5s to the end).
- A negative value counts backward from the end (e.g. <audio[-2.5s:]> means the last 2.5 seconds of that clip).
- The slicing semantics follow Python-style [start:end] conventions.
\end{lstlisting}
% Let the prompt continue naturally instead of forcing another page break.
\PromptGap
\begin{lstlisting}[style=paperprompt]
Guidelines:
- Base your answers only on information that is clearly present or can be directly inferred from the audio.
- Do NOT make up details that are not supported by the audio. Do not hallucinate sounds or speech that you did not hear.
- When describing audio content, always use the <audio[start:end]> notation to indicate when each element occurs in the audio.
- Be specific and factual about: speech content, speaker's emotional tone, speaking pace, voice characteristics, background sounds, acoustic environment.
- If the audio is noisy, ambiguous, or unclear, acknowledge the uncertainty rather than guessing.
\end{lstlisting}
\PromptGap
\begin{lstlisting}[style=paperprompt]
Your goal is to provide reliable, evidence-based answers grounded in the audio content only.
\end{lstlisting}
% END MESSAGE 3
\PromptDivider
\PromptRole{USER PROMPT}
% BEGIN MESSAGE 4
\begin{lstlisting}[style=paperprompt]
Listen to the provided audio <audio> carefully and create a chronological timeline of what you hear.
\end{lstlisting}
\PromptGap
\begin{lstlisting}[style=paperprompt]
MANDATORY SPEECH LABELING:
- EVERY speech segment MUST be labeled with <speech N> (N = 0, 1, 2, ... in chronological order)
- EVERY speaker MUST be labeled with <speaker N> (use the same N for the same speaker throughout)
- You MUST transcribe the spoken words as accurately as possible inside each speech segment
- If the audio contains ANY human voice, you MUST use <speech> type and provide transcription
- Do NOT describe speech as [sound_effect], [noise], or any other type. Speech = [speech] with <speech N> <speaker N> tags.
- Even brief utterances (hi, oh, hmm) must be tagged as <speech N>
\end{lstlisting}
\PromptGap
\begin{lstlisting}[style=paperprompt]
For each distinct audio element you detect, provide:
1. Time range using <audio[start:end]> notation
2. Type: [speech, music, sound_effect, silence, noise, other]
3. For speech: <speech N> <speaker N> followed by the exact transcription in quotes
   For non-speech: brief factual description
\end{lstlisting}
\PromptGap
\begin{lstlisting}[style=paperprompt]
Output format:
<audio[X.Xs:Y.Ys]> | [speech] | <speech 0> <speaker 0> "transcribed words here"
<audio[Y.Ys:Z.Zs]> | [music] | description of music
<audio[Z.Zs:W.Ws]> | [speech] | <speech 1> <speaker 1> "transcribed words here"
<audio[W.Ws:V.Vs]> | [sound_effect] | description of sound
...
\end{lstlisting}
\PromptGap
\begin{lstlisting}[style=paperprompt]
After the timeline, provide a scene summary:
\end{lstlisting}
\PromptGap
\begin{lstlisting}[style=paperprompt]
[Scene Summary]
<Summarize the overall scene: what is happening, where, who is involved, and the general atmosphere.>
- List each <speaker N> and describe their voice characteristics and, if identifiable, their likely role (e.g., "<speaker 0>: female, concerned tone, appears to be the mother; <speaker 1>: male, casual tone, appears to be a friend").
- Base this only on what you actually heard.
\end{lstlisting}
\PromptGap
\begin{lstlisting}[style=paperprompt]
Rules:
- Only describe sounds, speech, or music you are CERTAIN you heard
- Do NOT guess or invent any sounds you did not clearly hear
- If the audio is unclear or noisy, say so
- Use precise timestamps based on what you actually hear
- If uncertain about a boundary, use ~ (e.g., <audio[~5.0s:~8.0s]>)
\end{lstlisting}
% END MESSAGE 4
\end{paperpromptbox}

\begin{paperpromptbox}{promptAudio}{Targeted audio observation}{3}
\PromptRole{SYSTEM}
% BEGIN MESSAGE 5
\begin{lstlisting}[style=paperprompt]
You are an audio question answering model. Use the audio clip to answer the question.
\end{lstlisting}
% END MESSAGE 5
\PromptDivider
\PromptRole{USER CONTENT}
% BEGIN MESSAGE 6
\begin{lstlisting}[style=paperprompt]
audio_url: file://{audio_path}
text: {actor_ask_query}
\end{lstlisting}
% END MESSAGE 6
\end{paperpromptbox}

\begin{paperpromptbox}{promptExtractor}{Candidate-state extractor}{4}
\PromptRole{SYSTEM}
% BEGIN MESSAGE 7
\begin{lstlisting}[style=paperprompt]
You extract actor-visible candidate-state transitions. Return JSON only.
\end{lstlisting}
% END MESSAGE 7
\PromptDivider
\PromptRole{USER TEMPLATE}
% BEGIN MESSAGE 8
\begin{lstlisting}[style=paperprompt]
Reconstruct the candidate-answer state transition caused by one ASK in an audio reasoning trajectory. Use only the actor-visible question, options, audio description, reasoning, ASK, and Omni observation. Do not infer a hidden correct answer.
\end{lstlisting}
\PromptGap
\begin{lstlisting}[style=paperprompt]
[Task Question]
{question}
\end{lstlisting}
\PromptGap
\begin{lstlisting}[style=paperprompt]
[Candidate Answers]
{choices}
\end{lstlisting}
\PromptGap
\begin{lstlisting}[style=paperprompt]
[Initial Audio Description]
{initial_description}
\end{lstlisting}
\PromptGap
\begin{lstlisting}[style=paperprompt]
[Prior ASK and Actor-Visible Observation History]
{prior_history}
\end{lstlisting}
\PromptGap
\begin{lstlisting}[style=paperprompt]
[Actor Reasoning Before Current ASK]
{actor_reasoning}
\end{lstlisting}
\PromptGap
\begin{lstlisting}[style=paperprompt]
[Current ASK]
{current_ask}
\end{lstlisting}
\PromptGap
\begin{lstlisting}[style=paperprompt]
[Actual Actor-Visible Omni Observation]
{current_observation}
\end{lstlisting}
\PromptGap
\begin{lstlisting}[style=paperprompt]
Identify the options that were genuinely plausible before this ASK and those still plausible after the observation. An ASK is specific only when it targets a concrete, decision-relevant uncertainty that the observation could resolve. List only options whose status is supported by the visible trajectory. Eliminated distractors must be plausible before the ASK and no longer plausible after it.
\end{lstlisting}
\PromptGap
\begin{lstlisting}[style=paperprompt]
Return JSON only:
{
  "ask_is_specific": "<yes|no>",
  "plausible_before": ["<answer labels>"],
  "plausible_after": ["<answer labels>"],
  "eliminated_distractors": ["<answer labels>"],
  "reason": "<one or two sentence evidence-based explanation>"
}
\end{lstlisting}
% END MESSAGE 8
\end{paperpromptbox}

\endgroup % End the local prompt-appendix styles.

% BEGIN APPENDIX CASE EXAMPLES
\clearpage
\section{Qualitative Examples of Targeted Audio Interaction}

\label{app:qualitative-audio-cases}

\begingroup\raggedbottom

\noindent These recorded trajectories illustrate how targeted questions inform final answers. Observations are model reports; [...] marks omitted source text. Case titles are editorial.

\par

\begin{papercasebox}{caseAudio}{Case 1. Rhythm and timbre refine genre identification}

\CaseRole{caseAudio}{TASK QUESTION}

\CaseText{Which musical genre would this bass piece fit into?}

\CaseGap

\CaseRole{caseAudio}{CANDIDATE ANSWERS}

\CaseText{A. Blues\par B. Chamber music\par C. Funk\par D. Solo performance on bass}

\CaseDivider

\CaseRole{caseAudio}{INITIAL AUDIO DESCRIPTION}

\CaseText{<audio[0.0s:10.0s]> | [music] | Acoustic guitar playing a rhythmic, percussive melody.}
\CaseGap
\CaseText{[Scene Summary]\par The audio contains only instrumental music. There is no speech, sound effects, or background noise. The scene is a solo acoustic guitar performance.}

\CaseDivider

\CaseRole{caseActor}{ACTOR / REASONING BEFORE ASK}

\CaseText{The initial description mentions acoustic guitar with rhythmic, percussive melody, but doesn't describe bass elements specifically. I need to ask about what I can hear regarding bass characteristics like rhythm, texture, instrumentation, or musical style cues.}

\CaseDivider

\CaseRole{caseActor}{ACTOR / ASK 1}

\CaseText{<ask>What specific rhythmic patterns, bass line characteristics, or instrumental textures are audible in the piece?</ask>}

\CaseDivider

\CaseRole{caseAudio}{AUDIO MODEL / OBSERVATION 1}

\CaseText{The piece features a prominent, groovy, and syncopated bass line played on an electric bass guitar. The rhythm is built on a classic "slap" technique, where the player uses the thumb to strike the lower strings for a percussive "thump" and uses the fingers to pull the higher strings for a sharp "pop." This creates a distinct, funky, and driving rhythmic texture that is both melodic and percussive.}

\CaseDivider

\CaseRole{caseActor}{ACTOR / CONCLUSION}

\CaseText{The Omni Observation clearly describes a prominent electric bass line with syncopated, "slap" technique creating a funky, driving rhythmic texture. This aligns strongly with Funk music characteristics.}

\CaseResult{C}{C. Funk}

\end{papercasebox}

\clearpage

\begin{papercasebox}{caseAudio}{Case 2. Targeted listening reveals a background sound}

\CaseRole{caseAudio}{TASK QUESTION}

\CaseText{What non-musical sound can be heard in the recording?}

\CaseGap

\CaseRole{caseAudio}{CANDIDATE ANSWERS}

\CaseText{A. Clicking sounds\par B. String scrapes\par C. Footsteps in the background\par D. Amateur recording quality}

\CaseDivider

\CaseRole{caseAudio}{INITIAL AUDIO DESCRIPTION (EXCERPT)}

\CaseText{<audio[0.0s:10.0s]> | [music] | Soft, ambient music with a gentle, ethereal quality, featuring a female vocalist singing in a high, melodic tone.}
\CaseGap
\CaseText{[...]}
\CaseGap
\CaseText{There are no other discernible sounds or speakers.}

\CaseDivider

\CaseRole{caseActor}{ACTOR / REASONING BEFORE ASK}

\CaseText{The initial description mentions soft ambient music with a female vocalist and a speech fragment. I need to ask about specific non-musical sounds to distinguish between the options provided.}

\CaseDivider

\CaseRole{caseActor}{ACTOR / ASK 1}

\CaseText{<ask>Ask about specific non-musical sounds or ambient noises that occur throughout the recording to identify what is other than the music and vocal performance.</ask>}

\CaseDivider

\CaseRole{caseAudio}{AUDIO MODEL / OBSERVATION 1}

\CaseText{The most prominent non-musical sound is a rhythmic, mechanical clicking or tapping noise that occurs consistently throughout the entire recording, underneath the music and vocals.}

\CaseDivider

\CaseRole{caseActor}{ACTOR / CONCLUSION}

\CaseText{The Omni Observation clearly states there is a rhythmic, mechanical clicking or tapping noise throughout the recording, which corresponds to candidate A.}

\CaseResult{A}{A. Clicking sounds}

\end{papercasebox}

\clearpage

\begin{papercasebox}[2 ASKs]{caseEvidence}{Case 3. Deriving a disjunctive conclusion}

\CaseRole{caseEvidence}{TASK QUESTION}

\CaseText{What would be the most appropriate conclusion based on the given context?}

\CaseGap

\CaseRole{caseEvidence}{CANDIDATE ANSWERS}

\CaseText{A. I don't save money on gas or I take a taxi\par B. I save money on gas or I don't take a taxi\par C. I choose to take public transport and end up paying more for the ride\par D. I take a taxi and save money on gas}

\CaseDivider

\CaseRole{caseAudio}{INITIAL AUDIO DESCRIPTION (EXCERPT)}

\CaseText{<audio[0.0s:18.8s]> | [speech] | <speech 0> <speaker 0> "If I choose to take a public transport, I would be able to save money on gas. On the other hand, if I decide to take a taxi, I will end up paying more for the ride. It is known that at least one of the following statements is true. Either I opt for a public transport or I won't have to pay more for the ride. However, it is uncertain which one is true. It could be the case that only the first statement is true, or only the second statement is true, or perhaps both statements are true."}

\CaseDivider

\CaseRole{caseActor}{ACTOR / REASONING BEFORE ASK 1}

\CaseText{The speaker presents a logical puzzle about transportation choices. Let me analyze what is stated:}
\CaseGap
\CaseText{1. If public transport \ensuremath{\rightarrow} save money on gas\par 2. If taxi \ensuremath{\rightarrow} pay more for ride\par 3. At least one of these is true: (a) take public transport OR (b) won't pay more}
\CaseGap
\CaseText{The scenario seems to be about logical consistency around choices and costs.}

\CaseDivider

\CaseRole{caseActor}{ACTOR / ASK 1}

\CaseText{<ask>What specific logical constraints does the speaker state must hold true?</ask>}

\CaseDivider

\CaseRole{caseAudio}{AUDIO MODEL / OBSERVATION 1}

\CaseText{Based on the audio, the speaker states the following specific logical constraints:}
\CaseGap
\CaseText{1.  At least one statement is true: "It is known that at least one of the following statements is true."\par 2.  The statements are:\par - Statement 1: "Either I opt for a public transport or I won't have to pay more for the ride."\par - Statement 2: "I will end up paying more for the ride."\par 3.  The truth of the statements is uncertain: "However, it is uncertain which one is true. It could be the case that only the first statement is true, or only the second statement is true, or perhaps both statements are true."}
\CaseGap
\CaseText{In summary, the speaker states that the disjunction (logical OR) of the two statements must be true, but they do not specify which one(s) are true.}

\tcbbreak

\CaseRole{caseActor}{ACTOR / REASONING BEFORE ASK 2}

\CaseText{Let me formalize this:}
\CaseGap
\CaseText{Let P = take public transport\par Let S = save money on gas\par Let T = take taxi\par Let M = pay more for ride}
\CaseGap
\CaseText{From the audio:\par 1. If P then S (public transport saves gas)\par 2. If T then M (taxi means pay more)\par 3. At least one true: (P OR not M)\par 4. Uncertain which statement is true}
\CaseGap
\CaseText{The question asks what conclusion follows. Let me think about what logically follows from these premises.}
\CaseGap
\CaseText{If we know (P OR not M) is true, and we need to find what must follow...}
\CaseGap
\CaseText{Looking at the answer choices, they're about combinations of choices and costs.}

\CaseDivider

\CaseRole{caseActor}{ACTOR / ASK 2}

\CaseText{<ask>Does the audio indicate what statement(s) are actually true?</ask>}

\CaseDivider

\CaseRole{caseAudio}{AUDIO MODEL / OBSERVATION 2}

\CaseText{No, the audio does not indicate which statement(s) are actually true. It explicitly states that it is uncertain which one is true and that it could be the case that only the first statement is true, only the second statement is true, or both statements are true.}

\CaseDivider

\CaseRole{caseActor}{ACTOR / CONCLUSION}

\CaseText{So the constraint is: (P OR not M) is TRUE.}
\CaseGap
\CaseText{From the conditional statements:\par - P \ensuremath{\rightarrow} S (public transport \ensuremath{\rightarrow} save gas)\par - T \ensuremath{\rightarrow} M (taxi \ensuremath{\rightarrow} pay more)}
\CaseGap
\CaseText{Given (P OR not M), what follows?}
\CaseGap
\CaseText{If P is true, then S is true (save money on gas).\par If not M is true, then either P or not T (since T \ensuremath{\rightarrow} M, contrapositive gives not M \ensuremath{\rightarrow} not T).}
\CaseGap
\CaseText{Looking at answer B: "I save money on gas OR I don't take a taxi"}
\CaseGap
\CaseText{This would be S OR not T.}
\CaseGap
\CaseText{If we have (P OR not M):\par - Case 1: P is true \ensuremath{\rightarrow} S is true \ensuremath{\rightarrow} (S OR not T) is true\par - Case 2: not M is true \ensuremath{\rightarrow} From T \ensuremath{\rightarrow} M, if not M then not T \ensuremath{\rightarrow} (S OR not T) is true}
\CaseGap
\CaseText{Either way, (S OR not T) must be true.}
\CaseGap
\CaseText{So answer B follows logically.}

\CaseResult{B}{B. I save money on gas or I don't take a taxi}

\end{papercasebox}

\clearpage

\begin{papercasebox}{caseEvidence}{Case 4. Prosodic evidence supports sarcasm detection}

\CaseRole{caseEvidence}{TASK QUESTION}

\CaseText{Does the speaker express sarcasm or irony in the audio?}

\CaseGap

\CaseRole{caseEvidence}{CANDIDATE ANSWERS}

\CaseText{A. true\par D. false}

\CaseDivider

\CaseRole{caseAudio}{INITIAL AUDIO DESCRIPTION (EXCERPT)}

\CaseText{<audio[5.8s:8.3s]> | [speech] | <speech 2> <speaker 2> "Are we going to the drive-thru at Jack in the Box?"\par <audio[8.3s:9.5s]> | [noise] | audience laughter}
\CaseGap
\CaseText{[...]}
\CaseGap
\CaseText{- <speaker 2>: male, sarcastic and humorous tone, likely a character in a skit.}

\CaseDivider

\CaseRole{caseActor}{ACTOR / REASONING BEFORE ASK}

\CaseText{The scene summary notes that speaker 2 is described as having a "sarcastic and humorous tone." I should ask to confirm what specific evidence of tone or delivery supports this characterization.}

\CaseDivider

\CaseRole{caseActor}{ACTOR / ASK 1}

\CaseText{<ask>What specific auditory evidence supports the characterization of speaker 2 as having a sarcastic tone?</ask>}

\CaseDivider

\CaseRole{caseAudio}{AUDIO MODEL / OBSERVATION 1}

\CaseText{The specific auditory evidence supporting the characterization of speaker 2 as having a sarcastic tone is his exaggerated, mocking tone of voice.}
\CaseGap
\CaseText{Instead of a neutral or genuinely inquisitive tone, he delivers the line, "Are you going to the drive-through at Jack in the Box?" with a high-pitched, incredulous inflection. This delivery, combined with the immediate audience laughter that follows, indicates he is using the question rhetorically to express his disbelief and mock the idea that someone would spend an entire dinner discussing their research.}

\CaseDivider

\CaseRole{caseActor}{ACTOR / CONCLUSION}

\CaseText{The Omni Observation confirms that speaker 2 uses an exaggerated, mocking tone with incredulous inflection, suggesting irreverence or mockery. The audience laughter reinforces the humorous intent. This points toward sarcasm/irony in the delivery.}

\CaseResult{A}{A. true}

\end{papercasebox}

\clearpage

\begin{papercasebox}{caseEvidence}{Case 5. Vocal artifacts inform speech authenticity}

\CaseRole{caseEvidence}{TASK QUESTION}

\CaseText{Is the audio spoken by a real person or synthesized (fake)?}

\CaseGap

\CaseRole{caseEvidence}{CANDIDATE ANSWERS}

\CaseText{A. fake\par B. real}

\CaseDivider

\CaseRole{caseAudio}{INITIAL AUDIO DESCRIPTION}

\CaseText{<audio[0.0s:2.1s]> | [speech] | <speech 0> <speaker 0> "Is there something in particular that you want to eat?"}
\CaseGap
\CaseText{[Scene Summary]\par The audio contains a single speaker asking a question in a calm, conversational tone. The environment is quiet, with no background music or noise.}
\CaseGap
\CaseText{- <speaker 0>: male, calm tone, appears to be asking a question in a casual setting.}

\CaseDivider

\CaseRole{caseActor}{ACTOR / REASONING BEFORE ASK}

\CaseText{I should ask for specific evidence about vocal characteristics that would indicate whether the speech is real or synthesized.}

\CaseDivider

\CaseRole{caseActor}{ACTOR / ASK 1}

\CaseText{<ask>What specific vocal qualities, artifacts, or patterns appear in the audio that distinguish between natural human speech and synthesis?</ask>}

\CaseDivider

\CaseRole{caseAudio}{AUDIO MODEL / OBSERVATION 1 (EXCERPTS)}

\CaseText{1. Monotonous Pitch and Intonation: The pitch of the voice remains extremely flat and unchanging throughout the sentence.}
\CaseGap
\CaseText{[...]}
\CaseGap
\CaseText{2. Lack of Natural Rhythm and Prosody: The rhythm is perfectly regular and robotic.}
\CaseGap
\CaseText{[...]}
\CaseGap
\CaseText{3. Precise, Artificial Articulation: The pronunciation is unnaturally clear and consistent.}
\CaseGap
\CaseText{[...]}
\CaseGap
\CaseText{4. Absence of Breath Sounds and Micro-Phonations: Natural speech includes subtle breaths, lip smacks, and other physiological sounds. This voice is completely devoid of such background}

\CaseGap

{\rmfamily\itshape\fontsize{7.5}{9}\selectfont\color{caseEvidence}The source observation ends mid-sentence here.\par}

\CaseDivider

\CaseRole{caseActor}{ACTOR / CONCLUSION}

\CaseText{The Omni Observation provides clear evidence of artificial characteristics: monotonous pitch, lack of natural rhythm, precise artificial articulation, and absence of breath sounds. These are telling signs of synthesized speech rather than real human speech.}

\CaseResult{A}{A. fake}

\end{papercasebox}

\endgroup
% END APPENDIX CASE EXAMPLES

\end{document}